\documentclass[onecolumn,amsmath,amssymb,nofootinbib,12pt]{article}
\pdfoutput=1
\usepackage{jheppub}
\usepackage{amsfonts}
\usepackage{amsmath}
\allowdisplaybreaks[4]         
\usepackage{amssymb}   
\usepackage{euscript}       
\usepackage{color}    
\usepackage[header,title,page,titletoc]{appendix}

\newtheorem{theorem}{Theorem}
\newtheorem{corollary}{Corollary}

\newtheorem{defi}{Definition}
\newtheorem{conjecture}{Conjecture}

\usepackage{tensor}
\usepackage{braket}
\usepackage{float}
\newcommand{\be}{\begin{equation}}
\newcommand{\ee}{\end{equation}}

\newcommand{\bea}{\begin{eqnarray}}
\newcommand{\eea}{\end{eqnarray}}
\newcommand{\ba}{\begin{eqnarray}}
\newcommand{\ea}{\end{eqnarray}}

\newcommand{\beq}{\begin{equation}}
\newcommand{\eeq}{\end{equation}}
\newcommand{\beqa}{\begin{eqnarray}}
\newcommand{\eeqa}{\end{eqnarray}}
\newcommand{\beqar}{\begin{eqnarray*}}
\newcommand{\eeqar}{\end{eqnarray*}}

\newcommand{\eg}{{\it e.g.,}\ }
\newcommand{\ie}{{\it i.e.,}\ }

\newcommand{\req}[1]{(\ref{#1})}

\newcommand{\C}{\mathcal{C}}

\newcommand{\E}{\mathcal{E}}

\newcommand{\diff}{\mathrm{d}}

\renewcommand{\href}[2]{#2}

\arxivnumber{1906.00987}
\title{All higher-curvature gravities as\\ Generalized quasi-topological gravities}
\author[a]{Pablo Bueno,}
\author[b]{Pablo A. Cano,} 
\author[c]{Javier Moreno}
\author[b]{and \'Angel Murcia}

\affiliation[a]{Instituto Balseiro, Centro At\'omico Bariloche, \\
S. C. de Bariloche, R\'io Negro, R8402AGP, Argentina}
\affiliation[b]{Instituto de F\'isica Te\'orica UAM/CSIC,\\C/ Nicol\'as Cabrera, 13-15, C.U. Cantoblanco, 28049 Madrid, Spain\vspace{0.1cm}}
\affiliation[c]{Instituto de F\'isica, Pontificia Universidad Cat\'olica de Valpara\'iso,\\ Casilla 4059, Valpara\'iso, Chile.}

\vspace{0.2cm}
\emailAdd{pablo.bueno@cab.cnea.gov.ar}
\emailAdd{pablo.cano@uam.es}
\emailAdd{francisco.moreno.g@mail.pucv.cl} 
\emailAdd{angel.murcia@csic.es} 

\abstract{Generalized quasi-topological gravities (GQTGs) are higher-curvature extensions of Einstein gravity characterized by the existence of non-hairy generalizations of the Schwarzschild black hole which satisfy $g_{tt}g_{rr}=-1$, as well as for having second-order linearized equations around maximally symmetric backgrounds. In this paper we provide strong evidence that any gravitational effective action involving higher-curvature corrections is equivalent, via metric redefinitions, to some GQTG. In the case of theories involving invariants constructed from contractions of the Riemann tensor and the metric, we show this claim to be true as long as (at least) one non-trivial GQTG invariant exists at each order in curvature ---and extremely conclusive evidence suggests this is the case in general dimensions. When covariant derivatives of the Riemann tensor are included, the evidence provided is not as definitive, but we still prove the claim explicitly for all theories including up to eight derivatives of the metric as well as for terms involving arbitrary contractions of two covariant derivatives of the Riemann tensor and any number of Riemann tensors. Our results suggest that the physics of generic higher-curvature gravity black holes is captured by their GQTG counterparts, dramatically easier to characterize and universal.  As an example, we map the gravity sector of the Type-IIB string theory effective action in AdS$_5$ at order $\mathcal{O}({\alpha^{\prime}}^3)$ to a GQTG and show that the thermodynamic properties of black holes in both frames match.   }

\begin{document}
\maketitle
\flushbottom
\section{Introduction}
The gravitational effective action is expected to incorporate an infinite tower of higher-derivative corrections to the usual Einstein-Hilbert term. In particular, specific higher-curvature terms weighted by powers of $\alpha^{\prime}$ or $\ell_{\rm Planck}$ appear ---coupled to the rest of fundamental fields--- as stringy corrections in the different versions of String Theory \cite{Gross:1986mw,Gross:1986iv,Grisaru:1986vi,Grisaru:1986px,Bergshoeff:1989de,Green:2003an,Frolov:2001xr}.  Higher-curvature corrections are also naturally generated in the gravitational action when renormalizing quantum fields in curved spacetime ---see \eg \cite{Birrell:1982ix}--- including the graviton itself  \cite{DeWitt:1967yk,tHooft:1974toh,Goroff:1985th,vandeVen:1991gw,Solodukhin:2015ypa}.
From an Effective Field Theory (EFT) point of view, one can think of higher-derivative gravity as an effective description of some putative underlying UV-complete theory.  This approach requires the introduction of all possible terms compatible with the symmetries of the theory \cite{Weinberg:1995mt}. In the case of gravity, this means including all diffeomorphism-invariant higher-derivative operators available at each curvature order.

From a different perspective, particular classes of higher-derivative gravities presenting special properties have been often considered in various contexts. Sometimes, such theories can be used to test the genericness of Einstein's gravity predictions. For instance, Lovelock gravities \cite{Lovelock1,Lovelock:1971yv} provide natural generalizations of Einstein gravity in $D\geq 5$ whose second-order dynamics makes them particularly suitable for such comparisons.
The construction of higher-curvature theories which mimic and/or improve certain aspects of Einstein gravity has also been often explored ---\eg \cite{Stelle:1977ry,Stelle:1976gc,Lu,Tekin3,Maldacena:2011mk,Oliva:2011xu,Oliva:2012zs,Quasi2,PabloPablo,Lu:2015cqa,Tekin2,Aspects,Anastasiou:2016jix}. In some areas, like cosmology, the appearance of particular higher-derivative theories such as $f(R)$ gravities has become ubiquitous \cite{Sotiriou}.  Special mention deserves the role played by this class of theories within the holographic context \cite{Maldacena,Witten}. Higher-curvature gravities define toy models of strongly coupled CFTs inequivalent to Einstein gravity ---see \eg \cite{Hofman:2008ar,Camanho:2009vw,deBoer:2009pn,Buchel:2009sk,deBoer:2009gx,Grozdanov:2014kva,Andrade:2016rln,Konoplya:2017zwo,Quasi,Myers:2010jv,HoloRen,Hung:2014npa,deBoer:2011wk,Bianchi:2016xvf,Bueno:2018xqc} and references therein--- and they have been crucial in the discovery of certain universal results valid for general CFTs \cite{Myers:2010tj,Myers:2010xs,Mezei:2014zla,Bueno1,Bueno2} ---or to raise doubts on the possible universality of others \cite{Buchel:2004di,Kats:2007mq,Brigante:2007nu,Myers:2008yi,Cai:2008ph,Ge:2008ni}.

From an EFT point of view, there is an obvious issue with the classes of theories considered in the previous paragraph, namely, they involve specific combinations of higher-curvature invariants. On the other hand, in that context, one needs to take into account the possibility of performing field redefinitions ---we can redefine the metric tensor as $g_{ab}\rightarrow g_{ab}+K_{ab}$ for some symmetric tensor $K_{ab}$. Classically, the transformed field will, in general, have different properties from the original one. But from a more fundamental perspective, both fields should provide equivalent effective descriptions of the same underlying theory. Performing a field redefinition is equivalent to a change of variables, and this should leave the path integral, as well as scattering amplitudes and other observables, unaffected. Thus, even though the classical fields will be different, relevant physical properties are expected to be invariant under such redefinitions. 
In the gravitational context, this is the case of black hole's thermodynamic properties, such as temperature or entropy, which are invariant under field redefinitions of the metric \cite{Jacobson:1993vj} ---up to some subtleties we discuss below.  Thus, if our aim is to study black hole thermodynamics, we may work equivalently with the original theory or with the one resulting from a metric redefinition. 

A natural question one is led to ask is whether certain higher-curvature gravities possessing particularly interesting properties ---and which one would have naively considered ``fine-tuned'' from an EFT perspective--- may in fact be general enough so as to capture all terms appearing in the most general higher-derivative expansion once field-redefinitions are included in the game. In this paper we provide conclusive evidence that this is the case. In particular, we argue that any higher-curvature gravitational effective action constructed from arbitrary contractions of the metric and the Riemann tensor ---including the gravitational sector of any String Theory effective action--- can be mapped, via field redefinitions, to a theory of the so-called \emph{Generalized quasi-topological} class \cite{PabloPablo,Hennigar:2016gkm,PabloPablo2,Hennigar:2017ego,PabloPablo3,Ahmed:2017jod,PabloPablo4}. While we do not present a rigorous proof of this claim, we believe the evidence we provide is extremely compelling. In the case of terms involving covariant derivatives of the Riemann tensor, we are not able to provide the same degree of evidence, but we do prove the claim to be true for all effective actions of quartic order or lower as well as for terms consisting of two covariant derivatives of the Riemann tensor contracted in a generic way with an arbitrary number of Riemann tensors.

 As we review in detail in Section \ref{GQTGss}, Generalized quasi-topological gravities possess a number of remarkable properties such as the absence of additional modes besides the usual graviton propagating on maximally symmetric background, as well as the existence of non-hairy generalizations of the Schwarzschild-(AdS) black hole characterized by a single metric function $f(r)$ and whose thermodynamic properties can be easily accessed. Besides, the equation which determines such function is highly constrained and takes a universal form at each order in curvature: in $D=4$ all GQTG representatives contribute in exactly the same way to the equation ---see \eg \req{fequationn} below--- whereas in $D\geq5$ there are two classes of contributions ---the first involving only powers of $f(r)$ and the second involving up to two derivatives of $f(r)$.
 
 On the one hand, our results suggest a somewhat canonical way of writing a gravitational effective action, namely, as a series of non-trivial GQTG densities ---besides providing an additional motivation for the study of such class of theories. On the other, they suggest that the physics of black hole solutions to generic higher-curvature gravities which in principle appears to be very complicated ---\eg constructing the solutions themselves involves solving a different set of coupled differential equations (for the two functions appearing in the metric ansatz) for each possible theory considered--- can be actually mapped in all cases to the one of GQTG black holes, much more universal and easier to characterize.

A summary of our findings can be found next.

\subsection{Summary of results}\label{summy}
\begin{itemize}
\item[$\diamond$] Section \ref{GQTGss} contains a detailed review of GQTGs. This includes an account of their defining and most important features and explicit expressions for general GQTGs up to cubic order in arbitrary dimensions. Special emphasis is put on the high degree of simplicity and universality associated to the characterization of their static black hole solutions. 
\item[$\diamond$] Section \ref{riccis} starts with an explanation of the invariance of black hole thermodynamics under field redefinitions. Then, we explain how field redefinitions generically affect higher-derivative Lagrangians. In particular, we show that invariants involving Ricci curvatures ---or, more generally, those becoming a total derivative when evaluated on Ricci-flat backgrounds--- can always be removed from the action. 
\item[$\diamond$] In Section \ref{quadcub} we provide the explicit field redefinition which maps the most general quadratic and cubic gravities to GQTGs.  
\item[$\diamond$] In Section \ref{alGT} we reduce the problem of proving that any $\mathcal{L}(g_{ab},R_{abcd})$ gravity can be mapped to a GQTG to showing that at least one non-trivial GQTG exists at each order in curvature. Since extremely compelling evidence suggests that non-trivial GQTGs exist for any $D$ and at arbitrarily high orders in curvature, our result virtually proves that any higher-derivative effective action constructed from arbitrary contractions of the metric and the Riemann tensor can be mapped to a GQTG.
\item[$\diamond$] In Section \ref{covdiv} we turn to the more general case of higher-curvature densities involving covariant derivatives of the Riemann tensor. We show that all quartic gravities as well as densities constructed from an arbitrary number of Riemann tensors and two covariant derivatives can be mapped to GQTGs. In all cases, the  resulting GQTGs become equivalent to other GQTGs which do not involve any covariant derivative as long as static and spherically symmetric solutions are concerned.
\item[$\diamond$] In Section \ref{typeIIb} we use a field redefinition to map the gravity sector of the effective action of Type-IIB string theory truncated at (sub)leading order on AdS$_5\times \mathbb{S}^5$ to a GQTG. We compare the black hole solutions at leading order in the higher-curvature coupling in both frames and show that their thermodynamic properties match, as expected from our general analysis.
\item[$\diamond$] We conclude in Section \ref{discu} with some additional discussion and further reasonable conjectures. 
\end{itemize}

\section{Generalized quasi-topological gravities (GQTGs)}\label{GQTGss}
In recent years, an interesting new family of higher-curvature theories of gravity has been identified \cite{PabloPablo,Hennigar:2016gkm,PabloPablo2,Hennigar:2017ego,PabloPablo3,Ahmed:2017jod,PabloPablo4}. The action of these so-called \emph{Generalized quasi-topological gravities} (GQTGs) \cite{Hennigar:2017ego} can be written schematically as
\begin{equation}\label{priA}
S=\frac{1}{16\pi G}\int \diff^Dx \sqrt{|g|}\left[-2\Lambda+R+\sum_{n=2} \sum_{i_n} \ell^{2(n-1)}\mu_{i_n}^{(n)}\mathcal{R}_{i_n}^{(n)}\right]\, ,
\end{equation}
where $\ell$ is some length scale, $\mu^{(n)}_{i_n}$ are dimensionless couplings, and $\mathcal{R}^{(n)}_{i_n}$ are particular linear combinations of densities constructed in each case from contractions of $n$ Riemann tensors and the metric. The subindex $i_n$ refers to the number of independent GQTG invariants at each order $n$.

The technical requirement which makes a generic $\mathcal{L}(g^{ab},R_{abcd},\nabla_a R_{bcde},\dots)$ theory belong to the GQTG class is the following. 
Consider a general static and spherically symmetric ansatz (SSS),
\begin{equation}
\label{SSS}
\diff s^2_{\rm SSS}=-N(r)^2 f(r)\diff t^2+\frac{\diff r^2}{f(r)}+r^2\diff \Omega^2_{(D-2)}\, ,
\end{equation}
and let $L_{N,f}$ be the effective Lagrangian which results from evaluating $\sqrt{|g|}\mathcal{L}$ in \req{SSS}, namely
\begin{equation}\label{ansS}
L_{N,f}(r,f(r),N(r),f'(r),N'(r),\dots)= N(r) r^{D-2} \mathcal{L}|_{\rm SSS}\, ,
\end{equation}
(up to an irrelevant angular contribution). Also, let $L_f=L_{1,f}$, \ie the expression resulting from imposing $N=1$ in $L_{N,f}$. 

\begin{defi}\label{Def0}
We say that $\mathcal{L}(g^{ab},R_{abcd},\nabla_a R_{bcde},\dots)$ belongs to the GQTG family if the Euler-Lagrange equation of $L_f$ vanishes identically, \ie if
\begin{equation}\label{GQTGcond}
\frac{\partial L_f}{\partial f}-\frac{\diff}{\diff r} \frac{\partial L_f}{\partial f'}+\frac{\diff^2}{\diff r^2}  \frac{\partial L_f}{\partial f''}-\dots=0\, , \quad \forall \, \, f(r)\, .
\end{equation}
\end{defi}
The consequences of imposing \req{GQTGcond} have been explored quite extensively by now, and they can be summarized as follows.
\begin{enumerate}
\item When linearized around any maximally symmetric background, the equations of motion of GQTGs become second-order \ie they only propagate the usual massless and traceless graviton characteristic of Einstein gravity on such backgrounds \cite{PabloPablo,Hennigar:2016gkm,PabloPablo2,Hennigar:2017ego,PabloPablo3,Ahmed:2017jod,PabloPablo4}.\footnote{Note that higher-curvature gravities satisfying property ``1.'' ---and not necessarily the rest of properties appearing in the list, nor condition \req{GQTGcond}--- have been studied in several other papers, \eg \cite{Aspects,Tekin1,Tekin2,Tekin3,Love,Ghodsi:2017iee,Li:2017ncu,Li:2017txk,Li:2018drw,Li:2019auk,Lu:2019urr}.}

\item  They have a continuous and well-defined Einstein gravity limit, which corresponds to  setting $\mu^{(n)}_{i_n}\rightarrow 0$ for all $n$ and $i_n$.

\item They admit generalizations of the (asymptotically flat, de Sitter or Anti-de Sitter) Schwarzschild black hole  ---\ie solutions which reduce to it in the Einstein gravity limit--- characterized by a single function $f(r)$ \cite{Hennigar:2016gkm,PabloPablo2,Hennigar:2017ego,PabloPablo3,Ahmed:2017jod,PabloPablo4}. For them, $N(r)=1$ (or some other constant) in \req{SSS} and $g_{tt}g_{rr}=-1$.

\item The metric function $f(r)$ is determined from a differential equation of order $\leq 2$ ---which can be obtained from the Euler-Lagrange equation of $N(r)$ associated to the effective Lagrangian $L_{N,f}$ defined in \req{ansS}\footnote{Namely, the equation reads $\frac{\delta L_{N,f}}{\delta N}\big|_{ N=1}=0$, and it can be proven that it takes the form of a total derivative for any theory satisfying \req{GQTGcond}.}--- when the action does not include covariant derivatives of the Riemann tensor.\footnote{When it does, the differential equation would be of order $\leq 2m+2$, where $m$ is the number of covariant derivatives of the term with the greatest number of them.} Schematically, $\mathcal{E}[ r,f(r),f'(r),f''(r);\mu_{i_n}^{(n)}]=0$.
In that case, there are typically three situations:
\begin{itemize}
\item  The corresponding density does not contribute at all to the equation and then we call it ``trivial''.
\item The density contributes to the equation with an algebraic dependence on $f(r)$ ---namely, with terms involving powers of $f(r)$. This is the case of Quasi-topological \cite{Quasi2,Quasi,Dehghani:2011vu,Ahmed:2017jod,Cisterna:2017umf} and Lovelock \cite{Lovelock1,Lovelock:1971yv} terms. This kind of contributions only exist for $D\geq 5$.
\item The density contributes to the equation with terms containing up to two derivatives of $f(r)$. This is the case \eg of Einsteinian cubic gravity in $D=4$ \cite{PabloPablo,Hennigar:2016gkm,PabloPablo2}.
\end{itemize}
There is strong evidence that non-trivial GQTGs exist in any number of dimensions, including $D=4$, and for arbitrarily high orders of curvature. This evidence also suggests that the equation that determines $f(r)$ can only be modified in a single way at each order in curvature in $D=4$ and in two ways in $D\geq 5$. Namely, given a curvature order $n$, in $D=4$ there is a linear combination of parameters $\mu^{(n)}=\sum_{i_n} c_{i_n}\mu_{i_n}^{(n)}$
such that the contribution to the equation of $f(r)$ will only depend on $\mu^{(n)}$: as long as the equation of $f(r)$ is concerned, we can turn on and off as many densities as we want, provided at least one of them (corresponding to a nontrivial density) is nonzero at each order in curvature \cite{PabloPablo4}. The explicit form of the equation reads \cite{PabloPablo4}\footnote{In \cite{PabloPablo4}, examples of $n=3,4,5$ GQTG densities were constructed and used to guess the pattern for general $n$ in \req{fequationn}. Then, it was verified that the $n=6,7,8,9,10$ cases indeed agree with such pattern. Observe that the dependence on the curvature order is very simple, which strongly supports the claim that \req{fequationn} is valid for general $n\geq 11$ as well. In all cases, it was also verified that the ones appearing in \req{fequationn} are the only possible functional contributions from GQTGs to the equation of $f$ at each order $n$: the only effect of turning on an additional density at order $n$ is to  shift the value of $\mu^{(n)}$. }   
	\begin{align}\label{fequationn}
		(1-f)-\frac{2GM}{r}-\frac{\Lambda r^2}{3}-\sum_{n}\mu^{(n)} \ell^{2(n-1)} \frac{ {f'}^{(n-3)} }{r^{n-2} }  \left[\frac{f'^3}{n}+\frac{(n-3)f+2}{(n-1)r}f'^2 \right. \\  \notag \left.-\frac{2}{r^2}f(f-1)f'  -\frac{1}{r}f f''\left(f'r-2(f-1)\right)\right]=0
		\, ,
	\end{align}
where $M$ stands for the ADM mass of the solution \cite{Arnowitt:1960es,Arnowitt:1960zzc,Arnowitt:1961zz}.

 In $D\geq 5$ we can split the couplings in two sums of couplings. The first group of densities, belonging to the Quasi-topological subset, will modify the equation of $f(r)$ algebraically, whereas the second group will introduce derivatives of $f(r)$. The equation of $f(r)$ will only depend on a particular combination of couplings of each one of these groups \cite{Hennigar:2017ego,Ahmed:2017jod}. Schematically we have
 \begin{equation}\label{ecQG}
 \mathcal{E}_{\rm E}[r,f(r)]+\sum_n \left[ \mu^{(n)}_{\rm QT}\,  \mathcal{E}_n^{\rm QT}[r,f(r)^n]+ \mu^{(n)}_{\rm GQT}\,  \mathcal{E}_n^{\rm GQT}[r,f(r),f'(r),f''(r)]\right]=0\, ,
 \end{equation}
 where $ \mathcal{E}_{\rm E}[r,f(r)]$ is the Einstein gravity contribution 
 \begin{equation}
  \mathcal{E}_{\rm E}[r,f(r)]=(1-f)-\frac{16\pi GM}{(D-2)\Omega_{(D-2)}r^{D-3}}-\frac{2\Lambda r^2}{(D-1)(D-2)}\, ,
 \end{equation}
 where $\Omega_{(D-2)}=2\pi^{(D-1)/2}/\Gamma[(D-2)/2]$
 and $\mu^{(n)}_{\rm QT}$ and $ \mu^{(n)}_{\rm GQT}$ stand for the sums of all couplings corresponding to densities contributing algebraically and with derivatives of $f(r)$ to the equation respectively.  For planar and hyperbolic horizons, exactly the same story holds with small modifications in the corresponding equations for the metric function.

\item Both when the equation is algebraic and when it is differential of order $2$, given a fixed set of $\mu^{(n)}_{i_n}$, the equation admits a single black-hole solution representing a smooth deformation of Schwarzschild's one, which is completely characterized by its ADM energy.  For spherically symmetric configurations, the corresponding metric describes the exterior field of matter distributions \cite{PabloPablo3}.

\item The thermodynamic properties of black holes can be computed analytically by solving a system of algebraic equations without free parameters. At least in $D=4$, black holes typically become stable below certain mass, which substantially modifies their evaporation process \cite{PabloPablo4}.

\item A certain subset of GQTGs admit additional solutions of the Taub-NUT/Bolt class in even dimensions \cite{Bueno:2018uoy}. Similarly to black holes, these are also characterized by a single metric function and their thermodynamic properties can be computed analytically. 

\item A (generally) different subset of four-dimensional GQTGs also gives rise to second-order equations for the scale factor when evaluated on a Friedmann-Lema\^itre-Robertson-Walker ansatz, giving rise to a well-posed cosmological evolution \cite{Arciniega:2018fxj,Cisterna:2018tgx,Arciniega:2018tnn}. Remarkably, an inflationary period smoothly connected with late-time standard $\Lambda$-CDM evolution is naturally generated by the higher-curvature terms.

\end{enumerate}
In addition to this more or less structural properties, GQTGs have been considered in various  contexts, and many interesting additional properties and applications explored ---see \eg \cite{Dey:2016pei,Feng:2017tev,Hennigar:2017umz,Hennigar:2018hza,Bueno:2018xqc,Bueno:2018yzo,Poshteh:2018wqy,Mir:2019ecg,Mir:2019rik,Mehdizadeh:2019qvc,Erices:2019mkd,Emond:2019crr}.

At cubic order in curvature, the most general (nontrivial) GQTG can be written as
\begin{equation}\label{cubiGQTG}
S=\int \frac{\diff^Dx \sqrt{|g|}}{16\pi G} \left[-2\Lambda+R+\ell^2 \mu^{(2)}_1 \mathcal{X}_4+ \ell^4 \left(\mu^{(3)}_1 \mathcal{X}_{6}+\mu^{(3)}_2\mathcal{Z}_{D}+\mu^{(3)}_3\mathcal{S}_{D}\right)\right]\, ,
\end{equation}
where we used the notation of \req{priA} to denote the couplings. Here, $\mathcal{X}_4$ and $\mathcal{X}_6$ stand for the dimensionally-extended Euler quadratic and cubic densities, also known as Gauss-Bonnet and cubic Lovelock terms, respectively,
\begin{align}\label{GB}
\mathcal{X}_{4}=&+R^2-4R_{ab}R^{ab}+R_{abcd}R^{abcd}\, ,\\  \notag
\mathcal{X}_6=&-8\tensor{R}{_{a}^{c}_{b}^{d}}\tensor{R}{_{c}^{e}_{d}^{f}}\tensor{R}{_{e}^{a}_{f}^{b}}+4\tensor{R}{_{ab}^{cd}}\tensor{R}{_{cd}^{ef}}\tensor{R}{_{ef}^{ab}} -24\tensor{R}{_{abcd}}\tensor{R}{^{a bc}_{e}}R^{d e}\\ \label{lov3}
&+3\tensor{R}{_{abcd}}\tensor{R}{^{abcd}}R+24\tensor{R}{_{abcd}}\tensor{R}{^{ac}}\tensor{R}{^{bd}}
+16 R^{b}_{a} R_{b}^{c} R_{c}^{a}
-12R_{ab}R^{ab} R+R^3\, .
\end{align}
$\mathcal{X}_4$ is topological in $D=4$ and trivial for $D\leq 3$, while $\mathcal{X}_6$ is topological in $D=6$ and trivial for $D\leq 5$.
On the other hand, $\mathcal{Z}_D$ is the so-called Quasi-topological gravity density \cite{Quasi2,Quasi}
\begin{align}
\mathcal{Z}_D=&+{{{R_a}^b}_c{}^d} {{{R_b}^e}_d}^f {{{R_e}^a}_f}^c
               + \frac{1}{(2D - 3)(D - 4)} \left[ \frac{3(3D - 8)}{8} R_{a b c d} R^{a b c d} R \nonumber \right. \\ \label{ZD}
              & \left. - \frac{3(3D-4)}{2} {R_a}^c {R_c}^a R  - 3(D-2) R_{a c b d} {R^{a c b}}_e R^{d e} + 3D R_{a c b d} R^{a b} R^{c d} \right. \\ \nonumber
              & \left.
                + 6(D-2) {R_a}^c {R_c}^b {R_b}^a  + \frac{3D}{8} R^3 \right]\, .
\end{align}
Note that when only the above three terms are included in addition to the usual Einstein-Hibert action, the equation satisfied by the metric function $f(r)$ is algebraic ---which partially explains why they were identified before the last term, $\mathcal{S}_D$. We also stress that for $D\geq 6$, $\mathcal{Z}_D$ affects the equation of $f(r)$ in the same way as $\mathcal{X}_6$ does. For $D=5$, $\mathcal{X}_6$ is trivial, and the effect of $\mathcal{Z}_5$ is nontrivial ---from this perspective, we could have just omitted $\mathcal{X}_6$ from \req{cubiGQTG}. These observations are in agreement with our comments in ``4.'' above regarding the fact that at each order and for each $D$ there is a single way of modifying the equation of $f(r)$ algebraically (and another single way involving derivatives of $f(r)$ ---see below).

When $\mathcal{S}_D$ is included, the equation becomes differential of order $2$. The explicitly form of this density can be chosen to be \cite{Hennigar:2017ego}
\begin{equation}
\begin{aligned}
\mathcal{S}_D=&+14\tensor{R}{_{a}^{c}_{b}^{d}}\tensor{R}{_{c}^{e}_{d}^{f}}\tensor{R}{_{e}^{a}_{f}^{b}} +2\tensor{R}{_{a bcd}}\tensor{R}{^{a bc}_{e}}R^{d e}-\frac{(38-29 D+4 D^2)}{4(D-2)(2D-1)}\tensor{R}{_{abcd}}\tensor{R}{^{abcd}}R\\
&-\frac{2(-30+9D+4D^2)}{(D-2)(2D-1)}\tensor{R}{_{abcd}}\tensor{R}{^{ac}}\tensor{R}{^{bd}}
-\frac{4(66-35D+2D^2))}{3(D-2)(2D-1)}R^{b}_{a} R_{b}^{c} R_{c}^{a}\\
&+\frac{(34-21D+4D^2)}{(D-2)(2D-1)}R_{ab}R^{ab} R-\frac{(30-13D+4D^2)}{12(D-2)(2D-1)} R^3\, .
\end{aligned}
\label{SD}
\end{equation}
The explicit form of the equation of $f(r)$ corresponding to \req{cubiGQTG} can be found \eg in \cite{Hennigar:2017ego}.
In $D=4$, $\mathcal{S}_4$ is usually rewritten in terms of the so-called Einsteinian cubic gravity density,\footnote{The original construction of Einsteinian cubic gravity in \cite{PabloPablo} was based on the fact that it satisfies properties ``1.'' and  ``2.'' for general dimensions, and it does so in a way such that the relative coefficients appearing in its definition in \req{ECG} are the same for general $D$ ---just like for Lovelock theories. It was later realized that the four-dimensional version of the theory satisfies the rest of properties listed.} defined as \cite{PabloPablo}
\begin{equation}\label{ECG}
\mathcal{P}=12 R_{a\ b}^{\ c \ d}R_{c\ d}^{\ e \ f}R_{e\ f}^{\ a \ b}+R_{ab}^{cd}R_{cd}^{ef}R_{ef}^{ab}-12R_{abcd}R^{ac}R^{bd}+8R_{a}^{b}R_{b}^{c}R_{c}^{a}\, ,
\end{equation}
which was in fact the first GQTG identified beyond the Lovelock and Quasi-topological ones \cite{Hennigar:2016gkm,PabloPablo2}. Both densities are connected through \cite{Hennigar:2017ego}
\begin{equation}\label{s4ECG}
\mathcal{S}_4-\frac{1}{4}\mathcal{X}_6+4\mathcal{C}=\mathcal{P}\, ,
\end{equation}
where $\mathcal{C}$ is an example of a trivial GQTG, in the sense that it has no effect on the equation of $f(r)$, as its contribution to it vanishes identically. It is given by
\begin{equation}\label{s4C}
\mathcal{C}=\frac{1}{2}R_{a}^b R_b^a R-2R^{ac}R^{bd}R_{abcd}-\frac{1}{4}R R_{abcd}R^{abcd}+R^{de}R_{abcd}R^{abc}\,_e\, .
\end{equation}
Although in this paper we will not be particularly interested in trivial GQTGs, we emphasize that those terms are only trivial for SSS metrics, but they can ---and they do \cite{Arciniega:2018fxj,Cisterna:2018tgx,Arciniega:2018tnn}--- play an important role when other kinds of metrics are considered.

As we mention later on, the structure of GQTGs above described seems to extend to general dimensions and  arbitrary orders in curvature. So far, examples of GQTGs including covariant derivatives of the Riemann tensor have not appeared in the literature, but we are confident that they do exist as well ---see Sections \ref{covdiv} and \ref{discu} for discussions on the role played by invariants containing covariant derivatives of the Riemann tensor in our setup.

\section{Field redefinitions in higher-curvature gravities}\label{riccis}
In this section we explore some of the effects resulting from redefining the metric tensor on higher-curvature gravities. In Subsection \ref{osin}, we make some technical comments regarding metric redefinitions involving derivatives of the metric  itself and explain how on-shell actions evaluated on solutions related by metric redefinitions agree with each other. Then, in Subsection \ref{reduci}, we explain how higher-curvature densities involving Ricci curvatures ---or, more generally, densities which become a total derivative when evaluated on Ricc-flat metrics--- can be removed from the gravitational effective action by convenient metric redefinitions.
\subsection{On-shell action invariance}\label{osin}
Let us consider the most general metric-covariant theory of gravity\footnote{We also assume that parity is preserved so that we do not have to include terms containing the Levi-Civita symbol. Nevertheless, all results in this Section also apply when those terms are included.}
\begin{equation}\label{eq:generalhdg}
S[g_{ab}]=\int \diff^Dx\sqrt{|g|}\mathcal{L}\left(g^{ab}, R_{abcd}, \nabla_{e}R_{abcd},  \nabla_{e}\nabla_{f}R_{abcd},\ldots\right)\, .
\end{equation}
We are interested in determining how \req{eq:generalhdg} transforms under a redefinition of the metric tensor $g_{ab}$ of the form
\begin{equation}\label{eq:metricredef}
g_{ab}=\tilde g_{ab}+\tilde Q_{ab}\, ,
\end{equation}
where $\tilde Q_{ab}$ is a symmetric tensor constructed from the new metric $\tilde g_{ab}$. Ideally, we would like the field redefinition to be algebraic, so that the relation between $g_{ab}$ and $\tilde g_{ab}$ is functional. However, the most general tensor we can build using the metric without introducing higher derivatives is proportional to the metric itself. Hence, $\tilde Q_{ab}$ generically involves curvature tensors, and \req{eq:metricredef} is a differential relation. 
The action $\tilde S$ for the new metric $\tilde g_{ab}$ is simply obtained by substituting \req{eq:metricredef} in the original action, namely
\begin{equation}
\tilde S[\tilde g_{ab}]=S[\tilde g_{ab}+\tilde Q_{ab}]\, .
\end{equation}
Observe that since \req{eq:metricredef} involves derivatives of the metric, extremizing the action with respect to $\tilde g_{ab}$ is, in general, inequivalent from extremizing it with respect to $g_{ab}$.
Whenever $g_{ab}^{\rm sol}$ is a solution of the original theory, the relation \req{eq:metricredef} always produces a solution $\tilde g_{ab}^{\rm sol}$ of the transformed theory when we invert it. However, the converse is not true: there exist solutions of the equations of motion obtained from the variation with respect to $\tilde g_{ab}$ which do not produce a solution of the original theory when we apply the map \req{eq:metricredef}. The reason behind this is the presence of extra derivatives in the field redefinition. This increases the number of derivatives in the equations of motion derived from $\tilde S$, which introduces spurious solutions that need be discarded. This issue is further discussed in Appendix \ref{App:2}. Provided it is taken into account, both theories, $S$  and $\tilde S$, are equivalent.

Note that when we keep only the meaningful solutions ---\ie those which are related by \req{eq:metricredef}---  the corresponding on-shell actions match,
\begin{equation}\label{tiss}
\tilde{S}\left[\tilde{g}_{ab}^{\rm sol}\right]=S\left[g_{ab}^{\rm sol}\right]\, .
\end{equation}
Since, \eg black hole thermodynamics can be determined ---in the Euclidean path-integral approach  \cite{Gibbons:1976ue}--- by evaluating the on-shell action, this simple observation proves that black hole thermodynamics can be equivalently computed in both frames. 
The same conclusion can be reached \cite{Jacobson:1993vj} using Wald's formula \cite{Wald:1993nt} ---see  \cite{Exirifard:2006wa,Sen:2007qy,Liu:2008kt,Castro:2013pqa,Mozaffar:2016hmg} for additional discussions regarding this issue.\footnote{In order to prove this statement rigorously, it is necessary to assume some mild conditions on $\tilde Q_{ab}$, namely, its fall-off at infinity should be fast enough. All redefinitions  we will consider are well-behaved in this sense.}

Of particular interest for us will be situations in which both $g_{ab}^{\rm sol}$ and $\tilde g_{ab}^{\rm sol}$ represent static and spherically symmetric black holes. As argued in \cite{Jacobson:1993vj}, field redefinitions of the form \req{eq:metricredef} preserve both the asymptotic and horizon structures of $g_{ab}^{\rm sol}$, so they map black holes into black holes. Particularizing even more, from the following subsection on, we will consider higher-derivative theories controlled by small parameters and perturbative field redefinitions weighted by them.  Ultimately, one of the reasons for considering redefinitions mapping generic higher-derivative theories to GQTGs is the fact that the equations of motion of the latter on static and spherically symmetric configurations become particularly simple and universal. In this particular setup, \req{tiss} will relate the on-shell action corresponding to a certain generalization of the Schwarzschild-(A)dS black hole (continuously connected to it) for a given higher-derivative theory at leading order in the corresponding coupling to the on-shell action of the black hole solution corresponding to the  transformed GQTG. We will provide an explicit example of this match between on-shell actions in Section \ref{typeIIb}.

\subsection{Ricci curvatures and reducible densities}\label{reduci}
Let us now determine how the redefinition \req{eq:metricredef} changes the action \req{eq:generalhdg}. For that, we assume the redefinition to be perturbative, \ie we treat $\tilde Q_{ab}$ as a perturbation and we work at linear order. This is enough for our purposes, since, following the EFT approach, we will also expand the action in a perturbative series of higher-derivative terms. Observe that in this case the relation \req{eq:metricredef} can be inverted as
\begin{equation}\label{eq:metricredefinv}
\tilde g_{ab}=g_{ab}-Q_{ab}+\mathcal{O}(Q^2)\, ,
\end{equation}
where $Q_{ab}$ has the same expression as $\tilde Q_{ab}$ but replacing $\tilde g_{ab}\rightarrow g_{ab}$. 
Let us introduce the equations of motion of the original theory as 
\begin{equation}
\E_{ab}=\frac{1}{\sqrt{|g|}}\frac{\delta S}{\delta g^{ab}}\, .
\end{equation}
Then, at linear order in $\tilde Q_{ab}$, the transformed action $\tilde S$ reads
\begin{equation}\label{eq:tildeaction}
\tilde S=\int \diff^Dx\sqrt{|\tilde g|}\left[\tilde{\mathcal{L}}-\tilde{\E}_{ab}\tilde Q^{ab}+\mathcal{O}(Q^2)\right]\, .
\end{equation}
where the tildes denote evaluation on $\tilde g_{ab}$. Thus, the redefinition introduces a term in the action proportional to the equations of motion of the original theory
. Let us be more explicit about the form of the Lagrangian by expanding it as a sum over all possible higher-derivative terms
\begin{equation}\label{eq:faction}
S=\frac{1}{16\pi G}\int \diff^Dx\sqrt{|g|}\left[ R+\sum_{n=2}^{\infty} \ell^{2(n-1)}\mathcal{L}^{(n)}\right]\, ,
\end{equation}
where $\ell$ is a length scale and $\mathcal{L}^{(n)}$ represents the most general Lagrangian involving $2n$ derivatives of the metric. The explicit form of the invariants at orders $n=2$ and $n=3$ can be found below in \req{s4} and \req{s6} respectively. 
The number of terms grows very rapidly, and the $n=4$ Lagrangian already contains 92 terms \cite{0264-9381-9-5-003}.\footnote{Ref.~\cite{0264-9381-9-5-003} provides the number of \emph{linearly independent} invariants, but many of them differ by total derivative terms, which are irrelevant for the action. The number of relevant terms is, in general, much smaller ---yet quite large. For instance, besides the 3 quadratic densities and the 10 cubic densities which we include in \req{s4} and \req{s6}, \cite{0264-9381-9-5-003} adds $\nabla_a \nabla^a R$ to the former list, and 7 more terms of the form: $\nabla_a\nabla^a\nabla_b\nabla^b R$, $R\nabla_a \nabla^a R$, $\nabla^{a}\nabla^b R R_{ab}$, $R^{ab}\nabla_c \nabla^c R_{ab}$, $\nabla^a\nabla^b R^{cd}R_{cadb}$, $\nabla^a R^{bc}\nabla_{c}R_{ba}$, $\nabla^aR^{bcde}\nabla_a R_{bcde}$ to the latter. All these terms are either total derivatives or can be written in terms of the others plus total derivatives, so they can be discarded ---see \eg \cite{Oliva:2010zd,Quasi}. }

Let $\tilde Q_{ab}^{(k)}$ be a symmetric tensor containing $2k$ derivatives of the metric. Then, we perform the following field redefinition
\begin{equation}\label{eq:metricredefk}
g_{ab}=\tilde g_{ab}+\ell ^{2k}\tilde Q_{ab}^{(k)}\, .
\end{equation}
Then, the transformed action \req{eq:tildeaction} reads
\begin{equation}\label{eq:faction2}
\tilde S=\int \frac{\diff^Dx\sqrt{|\tilde g|}}{16\pi G} \left[\tilde R+\sum_{n=2}^{k} \ell^{2(n-1)}\tilde{\mathcal{L}}^{(n)}+ \ell^{2k}\left(\tilde{\mathcal{L}}^{(k+1)}-\tilde{R}^{ab}\hat Q_{ab}^{(k)}\right)+\sum_{n=k+2}^{\infty}  \ell^{2(n-1)}\tilde{\mathcal{L}}'^{(n)}\right]\, ,
\end{equation}
where all quantities are evaluated on $\tilde g_{ab}$, and\footnote{We have $\E^{ab}\tilde Q_{ab}^{(k)}=(R^{ab}-\frac{1}{2}g^{ab}R)\tilde Q_{ab}^{(k)}=R^{ab}\hat Q_{ab}^{(k)}$. 
}
\begin{equation}\label{qk}
\hat Q_{ab}^{(k)}=\tilde Q_{ab}^{(k)}-\frac{1}{2}\tilde g_{ab}\tilde Q^{(k)}\, ,\quad  \tilde Q^{(k)}=\tilde g^{ab}\tilde Q_{ab}^{(k)}\, .
\end{equation}
Hence, all terms containing up to $2k$ derivatives of the metric remain unaffected, while those with $2(k+1)$ derivatives receive a correction of the form $-\tilde{R}^{ab}\hat Q_{ab}^{(k)}$. The higher-order terms also get corrections which depend in a more complicated way on $\tilde Q_{ab}^{(k)}$. If the starting action already contained all possible terms, the net effect of these corrections is just to change the couplings in the Lagrangian. We denote these modified terms as $\tilde{\mathcal{L}}'^{(n)}$. 

From this, it is clear that performing this type of field redefinitions order by order, starting at $k=1$, we can remove all terms in the action which involve contractions of the Ricci tensor ---except, of course, the Einstein-Hilbert term. At each order, it suffices to choose $\tilde Q_{ab}^{(k)}$ in \req{eq:metricredefk} such that $\hat Q_{ab}^{(k)}$  equals  the tensorial structure which appears contracted with $R^{ab}$ in the corresponding density.
In other words, any term containing Ricci curvatures is meaningless from the EFT point of view,  and we are free to add or remove terms of that type.  From a different perspective, it has been argued ---\eg in \cite{Endlich:2017tqa}--- that if some higher-curvature correction controlled by $\ell^{2k}$ involves operators which vanish on the equations of motion produced by the lower-order action, the relevant physics is not affected at $\mathcal{O}(\ell^{2k})$, and we can just ignore it. For the gravitational effective action, this is equivalent to the possibility of removing all terms involving Ricci curvatures.



Observe that in \req{eq:faction} we (intentionally) did not include a cosmological constant. When we add it, the effect of the redefinition \req{eq:metricredefk} is 
\begin{equation}\label{eq:factioncosmo}
\begin{aligned}
\tilde S=&\int \frac{\diff^Dx\sqrt{|\tilde g|}}{16\pi G} \left[-2\Lambda+\tilde R+\sum_{n=2}^{k-1} \ell^{2(n-1)}\tilde{\mathcal{L}}^{(n)}+ \ell^{2(k-1)}\left(\tilde{\mathcal{L}}^{(k)}+\frac{2(\Lambda \ell^2)}{(D-2)}\hat Q^{(k)}\right)\right.\\
&\left.+\ell^{2k}\left(\tilde{\mathcal{L}}^{(k+1)}-\tilde{R}^{ab}\hat Q_{ab}^{(k)}\right)+\sum_{n=k+2}^{\infty} \ell^{2n}\tilde{\mathcal{L}}'^{(n)}\right]\, .
\end{aligned}
\end{equation}
Namely, not only the terms involving $2(k+1)$ derivatives of the metric get modified, those involving $2k$ derivatives also receive a correction.
This is a complication with respect to the case without cosmological constant. 
 If we remove terms involving Ricci curvatures at a given order, the field redefinition of the following order will introduce a correction of the form $\frac{2(\Lambda \ell^2)}{(D-2)}\hat Q^{(k)}$ which will generically include again terms involving Ricci curvatures. Hence, the process cannot be carried out order-by-order because all steps are coupled. If one wants to remove all the terms with Ricci curvature up to order $2k$, it is necessary to consider the most general field redefinition up to that order, \ie including all the terms $\tilde Q_{ab}^{(m)}$ of order $m\le k$ at the same time. Nevertheless, we stress that this is just a technical complication: finding the precise field redefinition that removes the corresponding Ricci curvature terms is more involved, but it can certainly be done. 

Motivated by the above analysis, let us close this section with a definition which will be useful in the remainder of the paper. 
\begin{defi}\label{Def1}
A curvature invariant is said to be ``reducible'' if it is a total derivative when evaluated on any Ricci-flat metric. The rest of them are said to be ``irreducible''.
\end{defi}
Note that this trivially contains  the case in which the invariant vanishes on Ricci-flat metrics. Intuitively, the irreducible terms correspond to those formed purely from contractions of the Riemann tensor, without explicit factors of Ricci curvature.   As we have explained, all reducible terms can be removed or introduced by using field redefinitions, whereas the irreducible ones cannot.
Therefore, the most general higher-derivative gravitational effective action is obtained by including all possible irreducible terms. Then, we are free to add as many reducible terms as we wish: these would simply correspond to different frame choices.

\section{All quadratic and cubic gravities as GQTGs }\label{quadcub}
In the absence of cosmological constant, the gravitational effective action can be written as a series of operators with an increasing number of derivatives of the metric:
\begin{equation}\label{eq:effectiveaction1}
S=\frac{1}{16\pi G}\int \diff^Dx \sqrt{|g|} R+ \sum_{n=2}^{\infty}\frac{\ell^{2n-2}}{16\pi G}S^{(2n)}\, .
\end{equation}
Again, $\ell$ is some length scale, and $S^{(2n)}$ is the most general action involving curvature invariants of order $n$. Ignoring total derivatives, the four- and six-derivative actions read
\begin{align}\label{s4}
S^{(4)}=\int \diff^Dx \sqrt{|g|}&\left[\alpha_1R^2+\alpha_2 R_{ab}R^{ab}+\alpha_3 R_{abcd}R^{abcd}\right]\, ,\\ \label{s6}
S^{(6)}=\int \diff^Dx\sqrt{|g|}&\left[\beta_1\tensor{R}{_{a}^{c}_{b}^{d}}\tensor{R}{_{c}^{e}_{d}^{f}}\tensor{R}{_{e}^{a}_{f}^{b}}+\beta_2 \tensor{R}{_{ab}^{cd}}\tensor{R}{_{cd}^{ef}}\tensor{R}{_{ef}^{ab}}+\beta_3 \tensor{R}{_{a bcd}}\tensor{R}{^{a bc}_{e}}R^{d e}\right.\\ \notag
&\left.+\beta_4\tensor{R}{_{abcd}}\tensor{R}{^{abcd}}R+\beta_5\tensor{R}{_{abcd}}\tensor{R}{^{ac}}\tensor{R}{^{bd}}+\beta_6R_{a}^{\ b}R_{b}^{\ c}R_{c}^{\ a}+\beta_7R_{ab }R^{ab }R\right.\\ \notag
&\left.+\beta_8R^3+\beta_9 \nabla_{d}R_{ab} \nabla^{d}R^{ab}+\beta_{10}\nabla_{a}R\nabla^{a}R\right]\, .
\end{align}
In the case of the four-derivative action, the Riemann-squared term can be traded by the Gauss-Bonnet density \req{GB},
so that the most general action reads\footnote{The coefficients $\alpha_i$ are not the same as in the previous action, but we prefer not to introduce additional unnecessary notation whenever possible.}
\begin{equation}
S^{(4)}=\int \diff^Dx \sqrt{|g|}\left[\alpha_1R^2+\alpha_2 R_{ab}R^{ab}+\alpha_3 \mathcal{X}_4\right]\, .
\end{equation}
Similarly to the quadratic case, we can trade two of the cubic invariants involving contractions of the Riemann tensor alone by  the cubic Lovelock density $\mathcal{X}_6$, defined in \req{lov3}, and one of the cubic Generalized Quasi-topological densities, $\mathcal{S}_D$, defined in  \req{SD}.
Therefore, $S^{(6)}$ can be alternatively written as
\begin{align}\notag 
S^{(6)}=&\int \diff^Dx\sqrt{| g|}\left[\beta_1\mathcal{X}_{6}+\beta_2 \mathcal{S}_{D}+\beta_3 \tensor{R}{_{a bcd}}\tensor{R}{^{a bc}_{e}}R^{d e}+\beta_4\tensor{R}{_{abcd}}\tensor{R}{^{abcd}}R+\beta_5\tensor{R}{_{abcd}}\tensor{R}{^{ac}}\tensor{R}{^{bd}}\right.\\ \label{sixs}
&\left.+\beta_6R_{a}^{\ b}R_{b}^{\ c}R_{c}^{\ a}+\beta_7R_{ab }R^{ab }R+\beta_8R^3+\beta_9 \nabla_{d}R_{ab} \nabla^{d}R^{ab}+\beta_{10}\nabla_{a}R\nabla^{a}R\right]\, .
\end{align}
Note that in $D\ge 5$, we can alternatively replace either $\mathcal{S}_{D}$ or $\mathcal{X}_6$ by the cubic Quasi-topological term $\mathcal{Z}_{D}$ defined in \req{ZD}. Also, in $D=4$ we can replace  $\mathcal{S}_{4}$ by the Einsteinian cubic gravity density \req{ECG} using \req{s4ECG}.
Regardless of these choices, we observe that in addition to the first two terms, belonging to the GQTG family, we are left  with a series of reducible terms which, as we have argued in the previous section, can be removed by convenient field redefinitions of the metric.  

The explicit redefinition which removes all terms but $\mathcal{X}_{4}$, $\mathcal{X}_{6}$  and $\mathcal{S}_{D}$ goes as follows.
First, in order to remove the $R^2$ and $R_{ab}R^{ab}$ terms, we perform 
\begin{equation}
g_{ab}=\tilde g_{ab}+\alpha_2\ell^2 \tilde R_{ab}-\frac{\ell^2 \tilde R}{D-2}\tilde g_{ab}(2\alpha_1+\alpha_2)\, .
\end{equation}
Then:
\begin{equation}
S^{(4)}\rightarrow \tilde{S}^{(4)}=\int \diff^Dx \sqrt{|\tilde g|}\alpha_3 \tilde{\mathcal{X}}_4\, .
\end{equation}
Now, this redefinition also affects the higher-order terms, but since we are starting from the most general theory, the only effect is to change the coefficients of these terms. In particular, for the six-derivative ones: $\beta_i\rightarrow\tilde\beta_{i}$. Then, the following redefinition of the metric
\begin{align}
&\tilde g_{ab}=\tilde{\tilde{g}}_{ab}+\ell^4\left[\tilde\beta_3\tensor{\tilde{\tilde R}}{_{a e cd}}\tensor{\tilde{\tilde R}}{_{b}^{e cd}}+\tilde\beta_5 \tilde{\tilde R}^{ef}\tilde{\tilde R}_{ae bf}+\tilde\beta_6 \tensor{\tilde{\tilde R}}{_{a}^{e}}\tilde{\tilde R}_{be}+\tilde\beta_{7}\tilde{\tilde R}\tilde{\tilde R}_{ab}-\tilde\beta_{9}\tilde{\tilde{\nabla}}^2\tilde{\tilde R}_{ab}\right.\\ \notag 
&\left. -\frac{\tilde{\tilde{g}}_{ab}}{D-2}\left(\tensor{\tilde{\tilde R}}{_{ef cd}}\tensor{\tilde{\tilde R}}{^{ef cd}}(\tilde\beta_3+2\tilde\beta_4)+\tilde{\tilde R}_{ef}\tilde{\tilde R}^{ef}(\tilde\beta_5+\tilde\beta_6)+\tilde{\tilde R}^2(\tilde\beta_7+2\tilde\beta_8)-\tilde{\tilde{\nabla}}^2 \tilde{\tilde R}(\tilde\beta_9-2\tilde\beta_{10})\right)
\right]\, ,
\end{align}
leaves the four-derivative terms unaffected, while cancelling all six-derivative terms that contain Ricci curvatures,
\begin{equation}
\tilde S^{(6)}\rightarrow \tilde{\tilde S}^{(6)}=\int \diff^Dx\sqrt{|\tilde{\tilde g}|}\left[\tilde{\beta_1}\tilde{\tilde{\mathcal{X}}}_{6}+\tilde\beta_2\tilde{\tilde{\mathcal{S}}}_{D}\right]\, .
\end{equation}
Hence, the most general action can be written, after all, as
\begin{equation}\label{eq:effectiveaction11}
\tilde{\tilde S}=\frac{1}{16\pi G}\int \diff^Dx \sqrt{|\tilde{\tilde g}|}\left[\tilde{\tilde{R}}+\ell^2 \alpha_3 \tilde{\tilde{\mathcal{X}}}_4+ \ell^4 \left( \tilde{\beta_1}\tilde{\tilde{\mathcal{X}}}_{6}+\tilde\beta_2\tilde{\tilde{\mathcal{S}}}_{D}\right)+\mathcal{O}(\ell^6)\right]\, ,
\end{equation}
which only contains GQTG terms, as anticipated ---compare with \req{cubiGQTG}. 
In $D=4$, the cubic Lovelock density vanishes identically and the Gauss-Bonnet term is topological, which leaves us with 
\begin{equation}
S=\frac{1}{16\pi G} \int \diff^4x \sqrt{|g|} \left[ R + \beta  \ell^4 \mathcal{P} + \mathcal{O}(\ell^6)\right]\, ,
\end{equation}
where we traded $\mathcal{S}_4$ by the ECG density $\mathcal{P}$ using \req{s4ECG} and we renamed the gravitational coupling. Hence, Einsteinian cubic gravity  \cite{PabloPablo} is (up to field redefinitions) the most general four-dimensional gravitational effective action we can write including up to six derivatives of the metric.\footnote{This is consistent with the result in \cite{Endlich:2017tqa}, where $\mathcal{P}$ appears traded by the density $\sim R_{ab}^{cd}R^{ab}_{ef}R^{ef}_{cd}$. That is also the kind of term which appears in the two-loop effective action of perturbative quantum gravity \cite{Goroff:1985th,vandeVen:1991gw}.}

\section{All $\mathcal{L}(g^{ab},R_{abcd})$ gravities as GQTGs}\label{alGT}
Let us now move on to a more general case, namely, general higher-curvature gravities constructed from arbitrary contractions of the metric and the Riemann tensor. In addition to the notion of ``reducible'' densities introduced in Section \ref{riccis}, it is convenient to  define here another concept:

\begin{defi}\label{Def2} 
We say that a curvature invariant $\mathcal{L}$ is  ``completable to a Generalized quasi-topological density'' (or just ``completable'' for short), if there exists a GQTG density $\mathcal{Q}$ such that $\mathcal{L}-\mathcal{Q}$ is reducible.
\end{defi}
In other words, $\mathcal{L}$ is completable if by adding reducible terms to it, we are able to obtain a GQTG term. Note that reducible terms are trivially completable to $0$.  Then, the question whether any higher-derivative gravity can be expressed as a sum of GQTG terms is equivalent to the following question: 
\emph{Are all irreducible densities completable to a GQTG?}
 We have just found that the answer is positive at least up to six-derivative terms. The reason is that there exist more independent GQTG densities than irreducible terms, which allowed us to ``complete'' all of them. In the case of the four-derivative terms, the only irreducible density is the Riemann-squared term, and this can be completed to the Gauss-Bonnet density. For the six-derivative terms, we saw that all terms containing derivatives of the Riemann tensor are reducible, and that the only irreducible terms are the two Riemann-cube contributions respectively controlled by $\beta_1$ and $\beta_2$ in \req{s6}. In general dimensions $D$ there are 3 GQTGs involving different combinations of these cubic terms, so they can always be completed. 

Observe that the problem of completing irreducible invariants depends on the number of spacetime dimensions. In lower dimensions, many of the densities  are not linearly independent, so the number of irreducible densities is significantly smaller, and this simplifies the problem of completing them to GQTGs. As a consequence, on general grounds we expect that if all irreducible invariants are completable for high enough $D$, they will also be completable for smaller $D$. For instance, going back to the six-derivative example, we find that the two cubic densities are independent when $D\ge 6$. In $D=4, 5$ only one of them is linearly independent, and in $D<4$ there is only Ricci curvature so all theories are reducible to Einstein gravity. On the other hand, the number of independent GQTGs in $D=4$ is four, whereas in $D>4$ there are only three of them. Therefore, in lower dimensions there are less irreducible terms and more ways to complete them to a GQTG theory. The lower the dimension, the easier the task. 

As we will see in a moment, the problem of completing all invariants constructed from an arbitrary contraction of metrics and $n$ Riemann tensors ---a number which grows very rapidly with $n$--- can be drastically simplified. In order to formulate this result, we will need the following somewhat surprising result:

\begin{theorem}[Deser, Ryzhov, 2005 \cite{Deser:2005pc}]\label{th1} 
When evaluated on a general static and spherically symmetric ansatz \req{SSS}, all possible contractions of $n$ Weyl tensors\footnote{Recall that the Weyl tensor is defined as
\begin{equation}
   W_{abcd}=R_{abcd}-\frac{2}{(D-2)}\left(g_{a[c}R_{d]b}-g_{b[c}R_{d]a}\right)+\frac{2}{(D-2)(D-1)}R g_{a[c}g_{d]b}\, .
\end{equation}} are proportional to each other. More precisely, let $(W^n)_i$ be one of the possible independent ways of contracting $n$ Weyl tensors, then for all $i$
\begin{equation}\label{Dser}
(W^n)_i|_{\rm SSS} = F(r)^n c_i  \, ,
\end{equation}
where $c_i$ is some constant which depends on the particular contraction, and $F(r)$ is an $i$-independent  function of $r$ given in terms of the functions appearing in the SSS ansatz \req{SSS}.  In other words, the ratio $[(W^n)_1/(W^n)_2]|_{\rm SSS}$ for any pair of contractions of $n$ Weyl tensors is a constant which does not depend on the radial coordinate $r$.
\end{theorem}

\noindent
\textit{Proof.} 
When evaluated on \req{SSS} 
 the Weyl tensor (with two covariant and two contravariant indices) takes the form
\begin{equation}\label{fack}
\left. \tensor{ W}{^{ab}_{cd}} \right|_{\rm SSS}=-2\chi(r) \frac{(D-3)}{(D-1)} \tensor{ w}{^{ab}_{cd}}\, ,
\end{equation}
where 
\begin{equation}
\chi(r)=\frac{(-2+2f-2r f'+r^2f'')}{2r^2}+\frac{N'}{2r N}(-2f +3r f')+\frac{f N''}{N}
\end{equation}
is a function which contains the full dependence on the radial coordinate. On the other hand, $\tensor{ w}{^{ab}_{cd}}$ is a $r$-independent tensorial structure which can be written as \cite{Deser:2005pc}
\begin{equation}\label{ww}
 \tensor{ w}{^{ab}_{cd}}=2\tau^{[a}_{[c} \rho^{b]}_{d]}-\frac{2}{(D-2)} \left(\tau^{[a}_{[c} \sigma^{b]}_{d]}+\rho^{[a}_{[c} \sigma^{b]}_{d]} \right)+\frac{2}{(D-2)(D-3)} \sigma^{[a}_{[c} \sigma^{b]}_{d]}\, ,
\end{equation}
where $\tau$, $\rho$ and $\sigma$ are orthogonal projectors defined as\footnote{Namely, they satisfy $\tau \tau=\tau$, $\rho \rho=\rho$, $\sigma \sigma=\sigma$, $\tau \rho=\tau\sigma=\rho\sigma=0$. Also, their traces read ${\rm Tr} \tau={ \rm Tr}\rho=1$, ${\rm Tr}\sigma=D-2$.}
\begin{equation}
\tau_{a}^{b}= \delta^0_{a} \delta^{b}_0 \, , \quad \rho_{a}^{b}= \delta^1_{a} \delta^{b}_1 \, , \quad  \sigma_{a}^{b}=\sum_{m=2}^{D} \delta^m_{a} \delta^{b}_m \, .
\end{equation}
The precise form of the projectors is not particularly relevant for our purposes. The important point is that any possible invariant $(W^n)_i$ constructed from the contraction of $n$ Weyl tensors will be given by
\begin{equation}\label{scs}
(W^n)_i|_{\rm SSS} =\left(-2 \chi(r) \frac{(D-3)}{(D-1)} \right)^n (w^n)_i\, ,
\end{equation}
where $ (w^n)_i$ stands for the constant resulting from the  contraction induced on the $w$ tensors, which we can identify with $c_i$ in  \req{Dser}. Therefore, $(W^n)_i|_{\rm SSS}$ takes the form \req{Dser} with $F(r)$ given by the function between brackets. $\, \square$



Now, we are ready to formulate one of the main results of the paper:
\begin{theorem}\label{th2} Let us consider the set of all irreducible curvature invariants of a given order which do not involve covariant derivatives of the curvature. If one of these invariants is completable to a GQTG and it does not vanish when evaluated on a static and spherically symmetric ansatz \req{SSS}, then all the invariants are completable.
\end{theorem}

\noindent
\textit{Proof.} 
Let the order of these invariants be $2n$ in derivatives of the metric, \ie $n$ in curvature. Since they are irreducible and they do not contain derivatives of the curvature, they are formed from contractions of a product of $n$ Riemann tensors. We can write schematically $\mathcal{L}_{i}=\left(\text{Riem}^n\right)_i$, where the subscript $i$ denotes again a specific way of contracting the indices. We can consider an alternative basis by replacing the Riemann tensor by the Weyl tensor in the expressions of these densities. Both ways of expressing these invariants are equivalent since they differ by terms containing Ricci curvatures, which are reducible. We denote the densities resulting from replacing $R_{abcd}\rightarrow W_{abcd}$ everywhere in the $\mathcal{L}_i$ by $\tilde{\mathcal{L}}_i=\left(W^n\right)_i$. Next, let us use the hypothesis of Theorem \ref{th2}, which consists in assuming that one of the densities, which we denote $\tilde{\mathcal{L}}_{i_0}$, is completable to a GQTG.  As we explained in Section \ref{GQTGss}, the condition that determines if a given density belongs to the GQTG class exclusively depends on the evaluation of the density on the general static and spherically symmetric (SSS) metric ansatz \req{SSS}, \ie on the way the corresponding density depends on the radial coordinate $r$. But from Theorem \ref{th1} we know that all order-$n$ invariants constructed from contractions of the Weyl tensor are proportional to each other when evaluated on \req{SSS}, in the sense that the dependence on the radial coordinate is identical for all $i$, and given by a fixed function ---which we called $F(r)^n$ in \req{Dser}. Then, since by assumption $\tilde{\mathcal{L}}_{i_0}\big|_{\rm SSS}\neq 0$, all invariants $\tilde{\mathcal{L}}_i$ are proportional to $\tilde{\mathcal{L}}_{i_0}$ when evaluated on SSS metrics. As a consequence, the fact that $\tilde{\mathcal{L}}_{i_0}\big|_{\rm SSS}$ is completable implies that all the rest of densities of order $n$ are, which concludes the proof. $\, \square$

The result can be reformulated as follows:
\begin{corollary} \label{coro1}
Let us consider the curvature invariants of a given order which do not involve derivatives of the curvature and let us assume that there exists one irreducible and non-trivial GQTG density formed from these invariants. Then, all  invariants are completable to GQTG densities. 
\end{corollary}
Recall that ``irreducible'' means that the density does not vanish on Ricci-flat metrics up to total derivatives whereas ``non-trivial'' means that it does not vanish for SSS metrics. We have compelling evidence that this type of GQTG theories exist at every order $n$ in curvature and for all $D$ ---and there are actually many of them, the number increasing rapidly with $n$. For instance, in $D=4$ examples of GQTG have been constructed explicitly up to $n=10$ in \cite{PabloPablo4,Arciniega:2018tnn}, where the general form of the equation satisfied by the metric function $f(r)$ in the SSS ansatz \req{SSS} was shown to have a simple dependence on the curvature order $n$ ---see \req{fequationn} above. Besides, in that case, the $n>3$ terms were constructed from products of a few $n=2$ and $n=3$ densities, and they already sufficed to produce examples of GQTG densities. Many more GQTGs could have been constructed had we not restricted the analysis to those building blocks (or even with different combinations of the same densities).   
Additional examples of GQTGs in $D>4$ and various curvature orders have also appeared in various papers \cite{Dehghani:2011vu,Hennigar:2017ego,Ahmed:2017jod,Cisterna:2017umf} ---\eg cubic and quartic densities up to $D=19$ have been explicitly verified to exist in \cite{Hennigar:2017ego} and \cite{Ahmed:2017jod}. For $D\rightarrow \infty$, the hypothesis of Corollary \ref{coro1} is guaranteed to be true for arbitrary $n$, since all Lovelock densities $\mathcal{X}_{2n}$ are irreducible and non-trivial in that case ---naturally, for finite $D$ the Lovelock family does not suffice, since all densities $\mathcal{X}_{2n}$ with $n\geq \lceil D/2 \rceil$ are either topological or trivial.

In sum, since: a) at a given order in curvature, we only need a single irreducible and non-trivial GQTG to exist in order for all invariants to be rewritable as GQTGs; b) numerous evidence strongly suggests that the number of independent irreducible and non-trivial GQTGs actually grows rapidly with the curvature order,
we are extremely confident to claim that, in any higher-curvature gravity, the terms which do not involve covariant derivatives of the curvature can always be written as a sum of GQTG densities by means of field redefinitions. 

Let us also note that the result above shows the existence of a field redefinition that takes the Lagrangian $\mathcal{L}(g^{ab},R_{abcd})$ to  a sum of GQTGs, but it does not guarantee unicity. Indeed, if at a given order one has several types of non-trivial GQTGs ---namely, Quasi-topological and proper GQTG--- it is possible to map the Lagrangian to a sum of terms whose equations for SSS metrics match the ones of the chosen theory (again, Quasi-topological or proper GQTG). More generally, the Lagrangian can be mapped to a combination of those terms. Note that this implies that Quasi-topological gravities and GQTGs are related by field redefinitions.\footnote{Imagine, for instance, that we start with a Quasi-topological density $\mathcal{Z}$ and a GQTG density $\mathcal{S}$ of certain order. Replacing all Riemann tensors by Weyl tensors gives rise to new densities $\tilde{\mathcal{Z}}=\mathcal{Z}+\text{RC}_{\mathcal{Z}}$ and $\tilde{\mathcal{S}}=\mathcal{S}+\text{RC}_{\mathcal{S}}$ where RC$_{\mathcal{S},\mathcal{Z}}$ are certain reducible densities involving Ricci curvatures. Now, from Theorem \ref{th1} we know that $\tilde{\mathcal{Z}}|_{\rm SSS}=c \tilde{\mathcal{S}}|_{\rm SSS}$, for some constant $c$. Then, it follows that $\mathcal{Z}= c(\mathcal{S}+\text{RC}_{\mathcal{S}})-\text{RC}_{\mathcal{Z}}+\mathcal{T}$, where $\mathcal{T}$ is a trivial GQTG density, \ie one such that $\mathcal{T}|_{\rm SSS}=0$. Naturally, $\mathcal{S}'\equiv c \mathcal{S}+\mathcal{T}$ is another GQTG density.  It follows that Quasi-topological densities can be mapped to GQTG densities of the same order ---and viceversa--- via field redefinitions, the mapping generically involving trivial GQTG densities (which play no role as far as the equations of SSS metrics are concerned).  } 

Before closing this section, let us mention that our conclusions also hold if one includes parity-breaking terms in the effective action, \ie those that involve the Levi-Civita symbol $\epsilon_{a_1\ldots a_D}$. In fact, all such terms vanish for spherically symmetric configurations, hence all of them trivially belong to the GQTG family.


\section{Terms involving covariant derivatives of the Riemann tensor}\label{covdiv}
In the previous section, we provided a strong argument in favor of the possibility that all $\mathcal{L}(g_{ab},R_{abcd})$ gravities can be either removed from the action or written as GQTGs using field redefinitions. Let us now see what happens with higher-curvature terms involving covariant derivatives of the Riemann tensor.
The role of these terms is less clear. In particular, they have not been used to construct GQTGs so far ---although, for what we know, this type of theories should exist as well. On the other hand, as we saw in Section \ref{quadcub}, up to six-order in derivatives all these terms are actually reducible. This is no longer the case at quartic order in curvature.

In order to gain some insight about the general behavior of this kind of terms, let us consider what happens at that order. There exist 26 independent quartic invariants which do not involve covariant derivatives of the Riemann tensor, namely ---see \eg \cite{0264-9381-9-5-003,Aspects,Ahmed:2017jod},
\begin{align}
\label{s8}
S^{(8)}=&\int \diff^Dx\sqrt{|g|}\left[\gamma_1 \tensor{R}{^{abcd}}\tensor{R}{_a^e_c^f}\tensor{R}{_e^g_b^h}\tensor{R}{_{fgdh}}+\gamma_2R^{abcd}\tensor{R}{_a^e_c^f} \tensor{R}{_{e}^g_f^h} \tensor{R}{_{bgch}}  \right. \\ \notag &+\gamma_3 R^{abcd} \tensor{R}{_{ab}^{ef}}\tensor{R}{_{c}^{g}_{e}^{h}} \tensor{R}{_{dgfh}}  +\gamma_4  R^{abcd} \tensor{R}{_{ab}^{ef}}  \tensor{R}{_{ce}^{gh}} \tensor{R}{_{dfgh}}
+\gamma_5  R^{abcd} \tensor{R}{_{ab}^{ef}}\tensor{R}{_{ef}^{gh}}R_{cdgh}
\\ \notag & +\gamma_{6}R^{abcd} \tensor{R}{_{abc}^{e}}\tensor{R}{_{fghd}}\tensor{R}{^{fgh}_e}  +\gamma_7 (R_{abcd}R^{abcd})^2+\gamma_8 R^{ab}\tensor{R}{^{cdef}}\tensor{R}{_c^g_{ea}}\tensor{R}{_{dgfb}}
\\ \notag & +\gamma_9R^{ab}\tensor{R}{^{cdef}}\tensor{R}{_{cd}^g_a}\tensor{R}{_{efgb}}   +\gamma_{10}R^{ab} \tensor{R}{_a^c_b^d} \tensor{R}{_{efgc}} \tensor{R}{^{efg}_{d}} +\gamma_{11} R \tensor{R}{_a^c_b^d} \tensor{R}{_c^e_d^f}\tensor{R}{_e^a_f^b}
\\ \notag &
+\gamma_{12}R R_{ab}^{cd} R_{cd}^{ef}R_{ef}^{ab}+\gamma_{13} R^{ab}R^{cd}\tensor{R}{^{e}_a^f_c}\tensor{R}{_{ebfd}}  +  \gamma_{14} R^{ab}R^{cd}\tensor{R}{^{e}_a^f_b}\tensor{R}{_{ecfd}}
\\ \notag &+\gamma_{15}R^{ab}R^{cd}\tensor{R}{^{ef}_{ac}}\tensor{R}{_{efbd}}+\gamma_{16}R^{ab}R_b^c \tensor{R}{^{def}_a}\tensor{R}{_{defc}}+\gamma_{17} R_{ef}R^{ef} R_{abcd}R^{abcd}
\\ \notag &+\gamma_{18}R R_{abcd} \tensor{R}{^{abc}_{e}}R^{de}+\gamma_{19}R^2 R_{abcd}R^{abcd}  +\gamma_{20} R^{ab}R_{acbd}R^{ec} R_{e}^d
\\ \notag &+\gamma_{21} R R_{abcd}R^{ac}R^{bd}+ \gamma_{22} R_a^b R_b^cR_c^dR_d^a+ \gamma_{23} \left(R_{ab}R^{ab} \right)^2 +\gamma_{24} R R_a^b R_b^cR_c^a \\ \notag & \left. +\gamma_{25} R^2 R_{ab}R^{ab}+ \gamma_{26} R^4\right]\, .
\end{align}
Of these, at most the first $7$ are irreducible ---this happens for $D>7$. Now, in \cite{Ahmed:2017jod} several non-trivial and irreducible GQTG theories were constructed using those invariants. Since by virtue of Corollary \ref{coro1} we only need one, this immediately implies that the 26 invariants can always be written as a sum of GQTGs using field redefinitions. Hence, just like in the quadratic and cubic cases, all quartic gravities of the form $\mathcal{L}(g_{ab},R_{abcd})$ can be written as GQTGs.

What about terms with covariant derivatives of the Riemann tensor? Looking at \cite{0264-9381-9-5-003}, we find five apparently irreducible terms of that kind, namely 
\begin{eqnarray}
\mathcal{L}_{1}&=&\tensor{R}{^{abcd}}\nabla_{b}\tensor{R}{^{efg}_{a}}\nabla_{d}\tensor{R}{_{efg c}}\, ,\\
\mathcal{L}_{2}&=&\tensor{R}{^{abcd}}\nabla_{c}\tensor{R}{^{efg}_{a}}\nabla_{d}\tensor{R}{_{efgb}}\, ,\\
\mathcal{L}_{3}&=&\tensor{R}{^{abcd}}\nabla^{g}\tensor{R}{^{e}_{a}^{f}_{c}}\nabla_{g}\tensor{R}{_{ebfd}}\, ,\\
\mathcal{L}_{4}&=&\tensor{R}{^{abcd}}\tensor{R}{_{a}^{efg}}\nabla_{d}\nabla_{g}\tensor{R}{_{becf}}\, ,\\
\mathcal{L}_{5}&=&\nabla_{e}\nabla_{f}\tensor{R}{_{abcd}}\nabla^{e}\nabla^{f}\tensor{R}{^{abcd}}\, .
\end{eqnarray}
However, a careful analysis ---using commutation of covariant derivatives, the symmetries of the Riemann tensor and the Bianchi identities\footnote{Recall that these read: $
R_{abcd}+R_{acdb}+R_{adbc}=0$ and  $\nabla_{e} R_{abcd}+\nabla_cR_{abde}+\nabla_d R_{abec}=0$ respectively.}--- reveals that all of them can be decomposed as a sum of total derivative terms plus quartic curvature terms (without covariant derivatives) plus terms with Ricci curvature (hence reducible). This is, they can be expressed (for each $i$) as
\begin{equation}\label{eq:Lred}
\mathcal{L}_{i}=\nabla_{a}J_{(i)}^{a}+\mathcal{Q}_{(i)}+R_{ab}F_{(i)}^{ab}\, ,
\end{equation}
for certain tensors $J_{(i)}^{a}$ and $F_{(i)}^{ab}$ and some quartic density $\mathcal{Q}_{(i)}$.
In order to illustrate this, let us show how $\mathcal{L}_1$ is reduced to an expression of the form \req{eq:Lred}. First, we have
\begin{align}
\mathcal{L}_{1}&=\tensor{R}{^{abcd}}\nabla_{b}\tensor{R}{^{efg}_{a}}\nabla_{d}\tensor{R}{_{efgc}}=\frac{1}{4}\tensor{R}{^{abcd}}\nabla^{g}\tensor{R}{^{ef}_{ab}}\nabla_{g}\tensor{R}{_{efcd}}\\ \notag
&=\frac{1}{4}\nabla_{g}\left(\tensor{R}{^{abcd}}\nabla^{g}\tensor{R}{^{ef}_{ab}}\tensor{R}{_{efcd}}\right)-\frac{1}{4}\tensor{R}{^{abcd}}\nabla^2\tensor{R}{^{ef}_{ab}}\tensor{R}{_{efcd}}-\frac{1}{4}\nabla_{g}\tensor{R}{^{abcd}}\nabla^{g}\tensor{R}{^{ef}_{ab}}\tensor{R}{_{efcd}}\, ,
\end{align}
where in the first equality we applied the differential Bianchi identity twice, and in the second we ``integrated by parts''.  Now we note that the last term in the second line is actually $-\mathcal{L}_{1}$, so we get
\begin{align}
\mathcal{L}_{1}&=\frac{1}{8}\nabla_{g}\left(\tensor{R}{^{abcd}}\nabla^{g}\tensor{R}{^{ef}_{ab}}\tensor{R}{_{efcd}}\right)-\frac{1}{8}\tensor{R}{^{abcd}}\nabla^g\nabla_g\tensor{R}{^{ef}_{ab}}\tensor{R}{_{efcd}}\, .
\end{align}
Then we are done, because the Laplacian of the Riemann tensor decomposes, using a schematic notation, as  $\nabla^2\text{Riem}=\nabla\nabla \text{Ricci}+\text{Riem}^2$,\footnote{Explicitly, one has \cite{Ricci}
\begin{align}
\nabla^e\nabla_eR_{abcd}=&+2\nabla_{[a|}\nabla_{c}R_{|b] d}+2\nabla_{[b|}\nabla_{d}R_{|a] c}-4\left[\tensor{R}{^p_a^q_b} R_{p[c| q |d]}+ \tensor{R}{^p_a^q_{[c|}} R_{pbq|d]}\right]\\ \notag &+g^{pq}\left[R_{qbcd}R_{pa}+R_{aqcd}R_{pb}�\right]\, .
\end{align}
} so we can indeed express $\mathcal{L}_{1}$ as in \req{eq:Lred}. 
Proceeding similarly with the rest of terms we arrive at the same conclusion. 

Since total derivatives are irrelevant for the action, and since we can remove all terms containing Ricci curvatures by means of field redefinitions, the terms with covariant derivatives of the Riemann tensor only change the coefficients of the quartic terms, which are already present in the action. Hence, from the point of view of effective field theory, these densities are meaningless and can be removed. In addition, we conclude that all eight-derivative terms can be recast as a sum of GQTGs by implementing field redefinitions.

Let us now turn to a more general case. Any higher-derivative gravity can be written as the span of all monomials formed from contractions of $\nabla_a$, $W_{abcd}$ and $R_{ab}$. Such a set can be written schematically as $\mathcal{A}=\cup_{q,n,r\in\mathbb{N}}\mathcal{A}_{q,n,r}$ where $\mathcal{A}_{q,n,r}=\{\nabla^{q}\times W^{n}\times {\rm Ric}^{r}\}$. Out of these subsets, the only ones susceptible of containing irreducible terms are $\mathcal{A}_{q,n,0}$, so the ultimate goal would be to prove that all elements in
\begin{equation}
\mathcal{I}_q=\bigcup_{n\in\mathbb{N}}\mathcal{A}_{q,n,0}
\end{equation}
are completable to a GQTG. First, let us note that these sets can be split according to the partitions of the number of covariant derivatives, $q$,
\begin{equation}
\mathcal{I}_q=\bigcup_{k=1}^{p(q)}\mathcal{I}^{P_k(q)}_q\, ,
\end{equation}
where $p(q)$ is the the number of partitions of $q$ and $P_k(q)$ denotes the $k$-th partition of $q$ (we assume  partitions to be ordered in some way). 
For instance, the first few cases are: $\mathcal{I}_0$, which is the set of monomials formed from general contractions of Weyl tensors; $\mathcal{I}_2$, which is the set of monomials formed from Weyl tensors and two covariant derivatives ---this can be in turn split as the union of $\mathcal{I}_2^{\{1,1\}}$ and $\mathcal{I}_2^{\{2\}}$: in the former set the two covariant derivatives act on two different Weyl tensors, while in the second the two derivatives act on the same Weyl; $\mathcal{I}_4$, which contains terms with four covariant derivatives and an arbitrary number of Weyl tensors ---this can be decomposed as $\mathcal{I}_4=\mathcal{I}_2^{\{1,1,1,1\}}\cup \mathcal{I}_2^{\{2,1,1\}}\cup  \mathcal{I}_2^{\{2,2\}}\cup  \mathcal{I}_2^{\{3,1\}}\cup  \mathcal{I}_2^{\{4\}}$.
Observe that not all subsets are independent. For example, we see that any term belonging to $\mathcal{I}_2^{\{2\}}$ can be written as a sum of terms in $\mathcal{I}_2^{\{1,1\}}$ upon integration by parts. For the same reason, for $q=4$ it is enough to keep the subsets $\mathcal{I}_2^{\{1,1,1,1\}}$, $\mathcal{I}_2^{\{2,1,1\}}$ and $\mathcal{I}_2^{\{2,2\}}$. 

We know that all terms in $\mathcal{I}_0$ can be completed to GQTGs, and the purpose of the remainder of this section is to show explicitly that all terms in $\mathcal{I}_2$ satisfy the same property.
We expect the trend to go on for all sets $\mathcal{I}_q$ but a general proof seems quite challenging ---not so much a case-by-case partial proof for the following $\mathcal{I}_{q\geq 4}$. 

As we have said, the only subset of $\mathcal{I}_{2}$ which needs to be considered is $\mathcal{I}_2^{\{1,1\}}$. Any term belonging to this subset can be written schematically as
\begin{equation}\label{eq:Inv2}
\mathcal{I}_2^{\{1,1\}}\ni\mathcal{R}_2^{\{1,1\}}=W^n \nabla W\nabla W\, ,
\end{equation}
for some value of $n$. We saw in \req{fack} that, when evaluated on a SSS metric the Weyl tensor has a very simple structure so that any scalar formed from it is proportional to the same quantity. In Appendix \ref{apricot} we show that any term of the form \req{eq:Inv2} can be written in turn as
\begin{equation}\label{eq:Inv3}
\left.\mathcal{R}_2^{\{1,1\}} \right|_{\rm SSS}=f(r)\chi^n\left(c_1(\chi')^2+c_2\frac{\chi\chi'}{r}+c_3\frac{\chi^2}{r^2}\right)\, ,
\end{equation}
where $\chi' =\diff \chi(r)/\diff  r$ and $c_{1,2,3}$ are constants. 
 Thus, there are at most three linearly independent terms in $\mathcal{I}_2^{\{1,1\}}$ when one considers SSS metrics. Hence, if we are able to find three independent terms in $\mathcal{I}_2^{\{1,1\}}$ which are completable to a GQTG, that will imply that all densities in $\mathcal{I}_2^{\{1,1\}}$ are completable. 
Three possible terms of that type are 
\begin{align}
\mathcal{W}^{\{1,1\}}_1=&\sum_{k=0}^n\nabla_{b}\tensor{W}{_{a_1a_2}^{a_3a_4}}\tensor{(W^{n-k})}{_{a_3 a_4}^{a_5 a_6}}\nabla^{b}\tensor{W}{_{a_5 a_6}^{a_7 a_8}} \tensor{(W^{n})}{_{a_7 a_8}^{a_1 a_2}}\, ,\\
\mathcal{W}^{\{1,1\}}_2=&\nabla_{b}\tensor{W}{^{bcd}_{a_1}}\nabla_{c}\tensor{W}{_{da_2}^{a_3a_4}}\tensor{W}{_{a_3a_4}^{a_5a_6}}\ldots \tensor{W}{_{a_{2n+1}a_{2n+2}}^{a_1a_2}}\, ,\\
\mathcal{W}^{\{1,1\}}_3=&\nabla_{b}\tensor{W}{^{bcde}}\nabla_{f}\tensor{W}{^{f}_{cde}}\tensor{W}{_{a_1a_2}^{a_3a_4}}\ldots \tensor{W}{_{a_{2n-1}a_{2n}}^{a_1a_2}} \, ,
\end{align}
where $\tensor{(W^{n})}{_{b c}^{d f}}$ denotes a $n$-Weyl product of the form $\tensor{W}{_{b c}^{a_1 a_2}} \tensor{W}{_{a_1 a_2}^{a_3 a_4}} \ldots \tensor{W}{_{a_{2n} a_{2n+1}}^{d f}}$. 
We can check that when evaluated on a SSS metric the previous terms are linearly independent. For instance, in $D=4$ we obtain the expressions
\begin{align}
\mathcal{W}^{\{1,1\}}_1=&\frac{3^{-n-2} 4 \left((-1)^n+2^{n+1}\right)f(r) (-\chi )^n \left( (n+1) r^2 \left(\chi '\right)^2+6 \chi ^2\right)}{r^2} \, ,\\
\mathcal{W}^{\{1,1\}}_2=&3^{-n-1} \left(2^n-(-1)^n\right)  f(r) (-\chi )^n \left(\frac{\chi '\chi}{r} +3 \frac{\chi^2}{r^2} \right)\, ,\\
\mathcal{W}^{\{1,1\}}_3=&f(r) \left( \chi '+3 \frac{\chi}{r} \right)^2 (-\chi)^n \frac{(2-(-1)^{n-1}2^{2-n})}{3} \, ,
\end{align}
which are linearly independent for any integer value of $n$. Hence, any term of the form \req{eq:Inv3} can be expressed a sum of these three combinations (the same conclusion holds for arbitrary $D$). Therefore all invariants in $\mathcal{I}_2^{\{1,1\}}$ can be expressed as a linear combination of these terms when evaluated on SSS metrics. This can be alternatively written as 
\begin{equation}\label{eq:Inv2b}
\mathcal{R}_2^{\{1,1\}}=C_1 \mathcal{W}^{\{1,1\}}_1+C_2\mathcal{W}^{\{1,1\}}_2+C_3\mathcal{W}^{\{1,1\}}_3+\ldots \, ,
\end{equation}
where the ellipsis denote terms that vanish on SSS metrics ---which are trivially completable to a GQTG. Now, it is easy to check that, by means of field redefinitions, the densities $\mathcal{W}^{\{1,1\}}_{1,2,3}$ are completable. Actually, both $\mathcal{W}^{\{1,1\}}_2$ and $\mathcal{W}^{\{1,1\}}_3$ are reducible because they are proportional to the divergence of Weyl tensor, which depends only on Ricci curvatures
\begin{equation}
\nabla_{c}\tensor{W}{^{c}_{abd}}=\frac{2(D-3)}{D-2}\left[\nabla_{[b}R_{d]a}-\frac{1}{2(D-1)}g_{a[d}\nabla_{b]}R\right]\, .
\end{equation}
On the other hand, $\mathcal{W}^{\{1,1\}}_1$ can be written as 
\begin{equation}
\begin{aligned}
\mathcal{W}^{\{1,1\}}_1&=\nabla_{b}\left(\nabla^{b}\tensor{W}{_{a_1a_2}^{a_3a_4}}\tensor{W}{_{a_3a_4}^{a_5a_6}}\tensor{W}{_{a_5a_6}^{a_7a_8}}\ldots\tensor{W}{_{a_{2n+3}a_{2n+4}}^{a_1a_2}}\right)\\
&-\nabla^2\tensor{W}{_{a_1a_2}^{a_3a_4}}\tensor{W}{_{a_3a_4}^{a_5a_6}}\tensor{W}{_{a_5a_6}^{a_7a_8}}\ldots\tensor{W}{_{a_{2n+3}a_{2n+4}}^{a_1a_2}}\, .
\end{aligned}
\end{equation}
Since the Laplacian of the Weyl tensor can be expressed as $\nabla^2\text{Weyl}=\nabla\nabla \text{Ricci}+\text{Riem}^2$, we conclude that, by means of field redefinitions, $\mathcal{W}^{\{1,1\}}_1$ can be reduced to a sum of terms without covariant derivatives. We know that those terms are completable, so the densities $\mathcal{W}^{\{1,1\}}_{1,2,3}$ and any other $\mathcal{R}_2^{\{1,1\}}$ are also completable. The result is actually stronger than that: since the densities $\mathcal{W}^{\{1,1\}}_{1,2,3}$ can be completed to a GQTG without covariant derivatives of the Riemann tensor, this implies that any other $\mathcal{R}_2^{\{1,1\}}$ can be completed to a GQTG which, when evaluated on a SSS metric, is equivalent to a GQTG without covariant derivatives. 

In sum, we have shown that, at least for densities including eight (or less) derivatives of the metric as well as for densities constructed from an arbitrary number of Riemann tensors and two covariant derivatives, all densities can be mapped to GQTGs. In all cases, those GQTGs become equivalent to GQTGs which do not involve covariant derivatives when evaluated on SSS metrics.  We postpone a discussion on the role of this kind of terms in ever more general situations to Section \ref{discu}. Before doing so, we wish to illustrate, for a particularly charismatic higher-derivative gravity action, how the mapping to a GQTG is done and how this preserves the thermodynamic properties of black hole solutions.

\section{Type-IIB effective action at $\mathcal{O}({\alpha^{\prime}}^3)$ as a GQTG}\label{typeIIb}
In this section we show how the gravitational sector of the Type-IIB String Theory effective action on AdS$_5\times \mathbb{S}^5$ truncated at (sub)leading order in $\alpha^{\prime}$ can be mapped to a generic quartic GQTG. Then we show that, in spite of the very different appearance of the equations of motion evaluated on a SSS ansatz  in both frames ---and therefore of the corresponding black hole metrics--- their thermodynamic properties exactly match, as expected.

The usual ten-dimensional Type-IIB supergravity action receives stringy corrections weighted by powers of  $\alpha'$. The first correction appears at ${\alpha'}^3$ order \cite{Gross:1986iv,Grisaru:1986px}, so schematically we have 
\begin{equation}\label{wtf}
   S_{\text{IIB}}=S_{\text{IIB}}^{(0)}+{\alpha'}^3 S_{\text{IIB}}^{(1)}+\ldots,
\end{equation}
where $S_{\text{IIB}}^{(0)}$ is the usual two-derivative supergravity action  \cite{Howe:1983sra}, and the dots stand for subleading corrections in  $\alpha'$. When the theory is considered in $\mathcal{A}_5\times\mathbb{S}^5$ where $\mathcal{A}_5$ is a negatively curved Einstein manifold, it is consistent to truncate all fields except the metric and it is possible to write an effective action for the five-dimensional metric \cite{Myers:2008yi,Buchel:2008ae,Galante:2013wta}. This is given by  \cite{Gubser:1998nz,Buchel:2004di}
\begin{equation}\label{acads5}
 S_{\text{IIB}_{{\mathcal{A}}_5 \times \mathbb{S}^5}}[g_{ab}]=\frac{1}{16\pi G}\int\diff^5 x\sqrt{|g|}\left[ R+\frac{12}{\ell^2}+\frac{\zeta(3)}{8}{\alpha'}^3 W^4 \right] \, ,
\end{equation}
where $W^4$ is a particular combination of contractions of four Weyl tensors given by
\begin{equation}
    W^4=\left(W_{abcd}W^{ebcf}+\frac{1}{2}W_{adbc}W^{efbc}\right)W\indices{^a^g_h_e}W\indices{_f_g^h^d}.
\end{equation}



As we mentioned in Section \ref{covdiv}, at quartic order in curvature, there are $26$ invariants involving contractions of the Riemann tensor of the metric ---see \req{s8}. The last $19$ densities involve explicit Ricci tensors, so they are reducible and we can use them to complete the Type-IIB effective action in \req{acads5} to GQTGs by means of field redefinitions. The structure of quartic GQTGs was completely characterized in \cite{Ahmed:2017jod}. As usual in $D\geq 5$, there exist three kinds of terms: those which belong to the Quasi-topological class (including the one previously constructed in \cite{Dehghani:2011vu} and the quartic Lovelock density $\mathcal{X}_8$) ---namely, their contribution to the equation which determines the metric function $f(r)$ when the SSS ansatz \req{SSS} is considered is algebraic; those which contribute with up to two derivatives to the equation of $f(r)$; those which do not contribute to the equation of $f(r)$ at all. As explained in Section \ref{GQTGss}, in spite of the degeneracy of GQTG densities, there are only two functional modifications of the equation of $f(r)$ at each curvature order, so when the full set of $n=4$ GQTG invariants is introduced, the different couplings only appear summed to each other in two groups in front of each kind of contribution to the equation as in \req{ecQG}.



Let us now consider the following metric redefinition
\begin{equation}
g_{ab}\rightarrow g_{ab}-\frac{\zeta(3)}{8}\alpha'^3\left(\hat C_{ab}-\frac{1}{3}\hat C g_{ab}\right)\, ,
\label{metredefi}
\end{equation}
where $\hat C_{ab}$ is some cubic-curvature rank-2 symmetric tensor and $\hat C=\hat C_{ab}g^{ab}$. 
The original action \req{acads5} is transformed to
\begin{equation}\label{tildeS}
\tilde S=\frac{1}{16\pi G}\int\diff^5 x\sqrt{|g|}\left[\frac{12}{\ell^2}+R+\frac{\zeta(3)\alpha'^3}{2\ell^2}\hat C+\frac{\zeta(3)}{8}{\alpha'}^3 \left(W^4+R^{ab}\hat{C}_{ab}\right) \right] \, ,
\end{equation}
up to subleading terms in $\alpha^{\prime}$.  Observe that the presence of the cosmological constant gives rise to the appearance of a cubic contribution.
The most general $\hat C_{ab}$ we can write is\footnote{We denote the different coefficients by $a_i$ and $b_i$. The $a_i$ correspond to terms which, when contracted with $R^{ab}$, produce a scalar numbered as in \req{s8}; the $b_i$ are used in cases in which there is a second term which produces the same scalar. }
\begin{align}
\hat C_{ab}=&a_8\tensor{R}{^{cdef}}\tensor{R}{_c^g_{ea}}\tensor{R}{_{dgfb}}
+a_9\tensor{R}{^{cdef}}\tensor{R}{_{cd}^g_a}\tensor{R}{_{efgb}}   +a_{10} \tensor{R}{_a^c_b^d} \tensor{R}{_{efgc}} \tensor{R}{^{efg}_{d}}\\ \notag &  +a_{11} g_{ab} \tensor{R}{_g^c_h^d} \tensor{R}{_c^e_d^f}\tensor{R}{_e^g_f^h}
+a_{12}g_{ab}\tensor{R}{_{gh}^{cd}}\tensor{R}{_{cd}^{ef}}\tensor{R}{_{ef}^{gh}}+a_{13} R^{cd}\tensor{R}{^{e}_a^f_c}\tensor{R}{_{ebfd}} \\ \notag &  +  a_{14} R^{cd}\tensor{R}{^{e}_a^f_b}\tensor{R}{_{ecfd}}
+a_{15}R^{cd}\tensor{R}{^{ef}_{ac}}\tensor{R}{_{efbd}}+a_{16}R_b^c \tensor{R}{^{def}_a}\tensor{R}{_{defc}}+a_{17} R_{ab}R_{cdef}R^{cdef}\\ \notag & 
+a_{18}g_{ab} R_{ghcd} \tensor{R}{^{ghc}_{e}}R^{de}+b_{18}R R_{ghca} \tensor{R}{^{ghc}_{b}}+a_{19}R_{ab}R R_{ghcd}R^{ghcd}+a_{20}R_{acbd}R^{ec} R_{e}^d \\ \notag &  +b_{20}R^{gh}R_{gahd} R_{b}^d
+a_{21} g_{ab} R_{ghcd}R^{gc}R^{hd}+b_{21} R R_{gacb}R^{gc}+ a_{22} \tensor{R}{_{a}^{c}}R_{{bc}}\\ \notag & + a_{23}R_{ab} R_{ef}R^{ef}+a_{24}g_{ab} \tensor{R}{_{c}^{d}} \tensor{R}{_{d}^{e}}\tensor{R}{_{e}^{c}}+b_{24} R  \tensor{R}{_{a}^{c}}R_{{bc}} +a_{25} g_{ab}R R_{ef}R^{ef} \\ \notag & +b_{25} R^2 R_{ab}+ a_{26} g_{ab}R^3\, .
\end{align}
Then we have
\begin{align}
R^{ab}\hat{C}_{ab}=&a_8 R^{ab}\tensor{R}{^{cdef}}\tensor{R}{_c^g_{ea}}\tensor{R}{_{dgfb}}
 +a_9R^{ab}\tensor{R}{^{cdef}}\tensor{R}{_{cd}^g_a}\tensor{R}{_{efgb}}   +a_{10}R^{ab} \tensor{R}{_a^c_b^d} \tensor{R}{_{efgc}} \tensor{R}{^{efg}_{d}}\\ \notag & +a_{11} R \tensor{R}{_a^c_b^d} \tensor{R}{_c^e_d^f}\tensor{R}{_e^a_f^b}
+a_{12}R\tensor{R}{_{ab}^{cd}}\tensor{R}{_{cd}^{ef}}\tensor{R}{_{ef}^{ab}}+a_{13} R^{ab}R^{cd}\tensor{R}{^{e}_a^f_c}\tensor{R}{_{ebfd}} \\ \notag &
 +  a_{14} R^{ab}R^{cd}\tensor{R}{^{e}_a^f_b}\tensor{R}{_{ecfd}}
+a_{15}R^{ab}R^{cd}\tensor{R}{^{ef}_{ac}}\tensor{R}{_{efbd}}+a_{16}R^{ab}R_b^c \tensor{R}{^{def}_a}\tensor{R}{_{defc}}\\ \notag &
+a_{17} R_{ef}R^{ef} R_{abcd}R^{abcd}
+(a_{18}+b_{18})R R_{abcd} \tensor{R}{^{abc}_{e}}R^{de}+a_{19}R^2 R_{abcd}R^{abcd} \\ \notag &
 +(a_{20}+b_{20}) R^{ab}R_{acbd}R^{ec} R_{e}^d
+(a_{21}+b_{21}) R R_{abcd}R^{ac}R^{bd}+ a_{22} R_a^b R_b^cR_c^dR_d^a\\ \notag &
+ a_{23} \left(R_{ab}R^{ab} \right)^2 +(a_{24}+b_{24}) R R_a^b R_b^cR_c^a +(a_{25}+b_{25}) R^2 R_{ab}R^{ab}+ a_{26} R^4\, ,
\end{align}
as well as
\begin{align}
\hat C=&(a_8+5a_{11})\tensor{R}{_{a}^{c}_{b}^{d}}\tensor{R}{_{c}^{e}_{d}^{f}}\tensor{R}{_{e}^{a}_{f}^{b}}+(a_9+5a_{12}) \tensor{R}{_{ab}^{cd}}\tensor{R}{_{cd}^{ef}}\tensor{R}{_{ef}^{ab}}\\ \notag
&+(a_{10}+a_{13}+a_{15}+a_{16}+5a_{18}) \tensor{R}{_{a bcd}}\tensor{R}{^{a bc}_{e}}R^{d e}+(a_{17}+b_{18}+5a_{19})\tensor{R}{_{abcd}}\tensor{R}{^{abcd}}R\\ \notag
&+(a_{14}+b_{20}+5a_{21})\tensor{R}{_{abcd}}\tensor{R}{^{ac}}\tensor{R}{^{bd}}+(a_{20}+a_{22}+5a_{24})R_{a}^{\ b}R_{b}^{\ c}R_{c}^{\ a}\\ \notag
&+(b_{21}+a_{23}+b_{24}+5a_{25})R_{ab }R^{ab }R+(b_{25}+5a_{26})R^3\, .
\end{align}
Imposing the terms $\hat C$ and $W^4+R^{ab}\hat{C}_{ab}$ to be of the GQTG type independently, we find the following constraints 
\begin{align}
a_{10}=&-\frac{43}{32}-\frac{13 \sigma }{32}-\frac{a_8}{10}-\frac{6a_9}{5}\, ,\\
a_{12}=&\frac{1}{16}+\frac{\sigma }{16}-\frac{a_8}{10}-\frac{a_9}{5}-\frac{a_{11}}{2}\, ,\\
a_{17}=&\frac{3451}{2880}+\frac{1241 \sigma }{2880}+\frac{3 a_8}{100}+\frac{3 a_9}{50}-\frac{7
   a_{13}}{40}-\frac{25 a_{14}}{72}-\frac{19 a_{15}}{180}-\frac{11 a_{16}}{45}\, ,\\
a_{18}=&-\frac{113}{640}-\frac{233 \sigma }{640}+\frac{31 a_8}{50}+\frac{6 a_9}{25}+3
   a_{11}-\frac{a_{13}}{5}-\frac{a_{15}}{5}-\frac{a_{16}}{5}\, ,\\
b_{18}=&-\frac{43}{640}-\frac{13 \sigma }{640}+\frac{7 a_8}{100}-\frac{9
   a_9}{25}-\frac{a_{13}}{5}-\frac{a_{15}}{5}-\frac{a_{16}}{5}\, ,\\
a_{19}=&-\frac{449}{2880}-\frac{17 \sigma }{1440}-\frac{19 a_8}{200}+\frac{3 a_9}{50}-\frac{3
   a_{11}}{8}+\frac{3 a_{13}}{40}+\frac{5 a_{14}}{72}+\frac{11 a_{15}}{180}+\frac{4 a_{16}}{45}\, ,\\
b_{20}=&-\frac{439}{144}-\frac{83 \sigma }{48}+\frac{3 a_8}{5}-\frac{24 a_9}{5}-2 a_{13}+\frac{4
   a_{14}}{3}-\frac{8 a_{15}}{3}-\frac{4 a_{16}}{3}-a_{20}\, ,\\
a_{21}=&\frac{1553}{1440}+\frac{391 \sigma }{480}-\frac{18 a_8}{25}+\frac{24 a_9}{25}-3 a_{11}+\frac{2
   a_{13}}{5}-\frac{7 a_{14}}{15}+\frac{8 a_{15}}{15}+\frac{4 a_{16}}{15}+\frac{a_{20}}{5}\, ,\\
b_{21}=&\frac{253}{288}+\frac{41 \sigma }{96}-\frac{a_8}{5}+\frac{6 a_9}{5}+\frac{3 a_{13}}{10}-\frac{11
   a_{14}}{30}+\frac{a_{15}}{3}+\frac{4 a_{16}}{15}-\frac{a_{20}}{5}\, ,\\
a_{23}=&-\frac{539}{810}-\frac{11 \sigma }{90}-\frac{4 a_8}{25}+\frac{56
   a_9}{75}+\frac{a_{13}}{3}+\frac{a_{14}}{90}+\frac{17 a_{15}}{45}+\frac{13 a_{16}}{45}-\frac{7
   a_{22}}{30}\, ,\\
a_{24}=&\frac{27}{64}+\frac{27 \sigma }{64}-\frac{2 a_8}{5}-2 a_{11}-\frac{a_{20}}{5}-\frac{a_{22}}{5}\, ,\\
b_{24}=&\frac{439}{960}+\frac{83 \sigma }{320}-\frac{7 a_8}{50}+\frac{18
   a_9}{25}+\frac{a_{13}}{5}-\frac{a_{14}}{5}+\frac{2 a_{15}}{5}+\frac{a_{20}}{5}-\frac{3
   a_{22}}{5}\, ,\\
a_{25}=&-\frac{599}{1296}-\frac{127 \sigma }{288}+\frac{2 a_8}{5}-\frac{8 a_9}{15}+\frac{3
   a_{11}}{2}-\frac{a_{13}}{6}+\frac{a_{14}}{9}-\frac{2
   a_{15}}{9}-\frac{a_{16}}{9}+\frac{a_{22}}{6}\, ,\\
b_{25}=&-\frac{317}{5184}-\frac{11 \sigma }{192}+\frac{a_8}{20}-\frac{7
   a_9}{30}-\frac{a_{13}}{24}+\frac{a_{14}}{24}-\frac{a_{15}}{12}+\frac{a_{22}}{6}\, ,\\
a_{26}=&\frac{1127}{25920}+\frac{41 \sigma }{960}-\frac{7 a_8}{200}+\frac{7
   a_9}{150}-\frac{a_{11}}{8}+\frac{a_{13}}{120}-\frac{a_{14}}{120}+\frac{a_{15}}{60}-\frac{a_{2
   2}}{30}\, .
\end{align}
We have 10 free parameters, which can be chosen to be $a_{8}$, $a_{9}$, $a_{10}$, $a_{11}$, $a_{13}$, $a_{14}$, $a_{15}$, $a_{16}$, $a_{20}$, $a_{22}$. However, we rewrote one of them, $a_{10}$, in terms of another constant that we called $\sigma$ ---this is convenient when studying black hole solutions as we show below. 

\subsection{Black hole solutions in the original frame}
Let us first study the black hole solutions of the Type-IIB action in the original frame \req{acads5}. We extend the spherical symmetry of \req{SSS} to planar and hyperbolic geometries as well, so that we search for solutions of the form
\begin{equation}\label{dsk}
\diff s^2=-N(r)^2f(r)\diff t^2+\frac{\diff r^2}{f(r)}+\frac{r^2}{\ell^2}\diff\Sigma_{k}^2\, ,\quad  \diff\Sigma_{k}^2=\begin{cases} \ell^2 \diff\Omega_{3}^2\, , \quad &\text{for}\quad k=1\, ,\\ \diff \vec{x}_{3}^2\, ,  \quad &\text{for}\quad k=0\, , \\  \ell^2 \diff \Xi_{3}^2 \, , \quad &\text{for}\quad k=-1\, .  \end{cases}
\end{equation}
The simplest way of computing the equations of motion is to use the reduced action method ---see \eg \cite{Palais:1979rca,Deser:2003up,PabloPablo2,PabloPablo3}.  After plugging the metric ansatz \req{dsk} into \req{acads5} and taking the corresponding functional derivatives with respect to $N(r)$ and $f(r)$, we proceed to solve the subsequent equations of motion perturbatively in $\alpha'$. Keeping only the leading $(\alpha')^3$ correction, we find the following expressions for $N(r)$ and $f(r)$:\footnote{Note that for general values of $\gamma$, the equations of motion of  \req{acads5} evaluated on the \req{dsk} ansatz become two fourth-order coupled differential equations for $N(r)$ and $f(r)$. This is in contrast with the GQTG frame, in which the corresponding equations of motion reduce to a single second-order equation for a single metric function ---see \req{gaga} below--- for general $\gamma$.}
\begin{equation}
\begin{split}\label{fN}
f(r)&=k+\frac{r^2}{\ell^2}\left[1-\frac{\omega^4}{r^4}+  \gamma\left(\frac{360 \omega^{12} }{r^{12}}+\frac{320 \omega^{12} k\ell^2 }{r^{14}}-\frac{285 \omega^{16}}{r^{16}}\right)\right]\, ,\\
N(r)&=N_k\left(1-\gamma\frac{120\omega^{12}}{r^{12}}\right)\, ,
\end{split}
\end{equation}
where we introduced $\gamma=\zeta(3)\alpha'^3/(8\ell^6)$ and where $N_k$ and $\omega^4$ are integration constants. In particular, $\omega^4$ defined in this way is proportional to the total energy of the solutions. In the $k=-1$ case, the expressions in \req{fN} can be seen to agree with those appearing in \cite{Galante:2013wta}, but one should take into account that the integration constants have been chosen differently. 
Now, the temperature $T$ of any black hole solution of the type considered is given by
\begin{equation}
T=\frac{N(r_h) f'(r_h)}{4\pi}\, ,
\end{equation}
where $r_h \equiv\max \{r_i | f(r_i)=0 \}$ is the value of the radial coordinate at which the event horizon is located. As a function of $r_h$, the temperature and the parameter $\omega^4$ read 
\begin{align}\label{TrhIIB}
T&=\frac{N_k}{4\pi}\left[\frac{2  k}{r_h}+\frac{4 r_h}{\ell^2}+\gamma\left (\frac{60   r_h}{\ell^2}-\frac{20   k^4 \ell^6}{r_h^7}+\frac{120    k^2 \ell^2}{r_h^3}+\frac{160    k}{r_h} \right ) \right]\, ,\\
\omega^4&=r_h^4+k\ell^2r_h^2+5\gamma  \left(r_h^2+k \ell^2\right)^3 \left(\frac{15}{ r_h^2}+\frac{7 k \ell^2}{r_h^4}\right)\, ,
\end{align}
where again we are working perturbatively in $\gamma$. 
Let us now compute the on-shell action of these solutions in order to determine their free energy, from which we can obtain the rest of relevant thermodynamic quantities. In order to do this, we need to include an appropriate generalized Gibbons-Hawking-York term \cite{Gibbons:1976ue} as well as counterterms for the action \req{acads5}. To the best of our knowledge, specific boundary terms have not been constructed for this theory. However, we can use the effective boundary terms introduced in \cite{Bueno:2018xqc}. In that reference, it was argued that for theories with second-order linearized equations of motion around maximally symmetric backgrounds, one can write an effective boundary term that works for asymptotically AdS solutions. The prescription is that the same GHY term and counterterms that appear for Einstein gravity must be multiplied by an overall constant, which in the holographic context is identified with the universal contribution to the entanglement entropy across a spherical region, $a^*$ ---see \eg \cite{Myers:2010tj,CHM}. In the case of the theory \req{acads5}, the condition of second-order linearized equations is satisfied ---in fact, the Weyl$^4$ term does not contribute to the linearized equations at all--- and the charge $a^*$ coincides with the Einstein gravity one. Therefore, we can use directly the same boundary terms and counterterms as for Einstein gravity \cite{Henningson:1998gx,Balasubramanian:1999re,Emparan:1999pm}, and the Euclidean action reads
\begin{align}\label{S5E}
 S^{\rm E}_{\text{IIB}_{{\mathcal{A}}_5 \times \mathbb{S}^5}}=-\int_{\mathcal{M}} \frac{\diff^5 x\sqrt{|g|}}{16\pi G}\left[R+\frac{12}{\ell^2}+\gamma \ell^6 W^4 \right]- \int_{\partial \mathcal{M}} \frac{\diff^{4}x\sqrt{h}}{8\pi G}\bigg[K-\frac{3}{\ell}-\frac{\ell}{4}\mathcal{R}\Bigg]\, .
\end{align}
The computation is more or less straightforward, and we get the result
\begin{equation}
S^{\rm E}_{\text{IIB}_{{\mathcal{A}}_5 \times \mathbb{S}^5}}=\frac{\beta N_k V_k}{16\pi G\ell^5 }\left[\frac{3k^2\ell^4}{4}+k\ell^2r_h^2-r_h^4+\frac{5\gamma}{r_h^4}\left(k \ell^2-15 r_h^2\right) \left(k \ell^2+r_h^2\right)^3 \right]\, ,
\end{equation}
where $V_k$ is the dimensionful volume of the transverse space (for instance, $V_1=2\pi^2\ell^3$) and $\beta$ is the inverse temperature, corresponding to the Euclidean time periodicity. 
When we express this result in terms of the black hole temperature we get 
\begin{equation}\label{SEIIB}
\begin{split}
S^{\rm E}_{\text{IIB}_{{\mathcal{A}}_5 \times \mathbb{S}^5}}=&\frac{V_k}{32 G} \bigg[3k x-x^3\mp\left(x^2-2k\right)^{3/2} \\&-\frac{15}{2}\gamma \left (\frac{k^2}{x}-28 k x+34 x^3\pm(6k-30 x^2) \sqrt{x^2-2k} \right) \bigg]\, ,
\end{split}
\end{equation}
 where we have introduced the notation $x=\pi \ell T/N_k$. 


\subsection{Black hole solutions in the GQTG frame}
Let us now compare this result with the one corresponding to the transformed GQTG frame  \req{tildeS}. 
This theory possesses black hole solutions characterized by a single function, namely, of the form 
\begin{equation}
\diff s^2=-N_k^2 f(r)\diff t^2+\frac{\diff r^2}{f(r)}+\frac{r^2}{\ell^2}\diff\Sigma_{k}^2\, ,\quad  \diff\Sigma_{k}^2=\begin{cases} \ell^2 \diff\Omega_{3}^2\, , \quad &\text{for}\quad k=1\, ,\\ \diff \vec{x}_{3}^2\, ,  \quad &\text{for}\quad k=0\, , \\  \ell^2 \diff \Xi_{3}^2 \, , \quad &\text{for}\quad k=-1\, ,  \end{cases}
\end{equation}
where $N_k$ is now a constant.
It is convenient to write  $f=k+g(r) r^2/\ell^2$. Then, the equation which determines the metric function reads 
\begin{align}  \notag \label{gaga}
g-\left(1-\frac{\omega ^4}{r^4}\right)=&-\frac{5\zeta(3){\alpha'}^3}{2048\ell^6}  g' \Big[-8 \left(2 r^3 (1+\sigma )+3 g r^3 (1+2
   \sigma )+2 k \ell^2 r (3+5 \sigma )\right) g'^2\\  \notag
   &
   +3 r^4 (1+2 \sigma )
  g'^3-48 (-1+g) r \left(k \ell^2+g r^2\right) (1+\sigma ) g''\\   &-12 g' \big(10 g^2 r^2
   (1+\sigma )+k \ell^2 \left(-14 (1+\sigma )+r^2 (1+2 \sigma ) g''\right)\\ \notag
   &+g \left(2 \left(7 k
   \ell^2-5 r^2\right) (1+\sigma )+r^4 (1+2 \sigma ) g''\right)\big)\Big]\, ,
\end{align}
where again $\omega^4$ is an integration constant related to the ADM energy of the solution.
Observe that, while there are a lot of independent free couplings, they all affect the equation of $g(r)$ in a very universal way controlled by the combination of coefficients given by $\sigma$. 
Solving perturbatively the above equation one is left with
\begin{equation}
\begin{split}
f(r)&=k+\frac{r^2}{\ell^2}\left[1-\frac{\omega^4}{r^4}+  \gamma\left(\frac{5 (5-13 \sigma ) \omega ^{12}}{2 r^{12}}+\frac{5 k \ell^2 (3-11 \sigma ) \omega
   ^{12}}{2 r^{14}}+\frac{15 (-1+3 \sigma ) \omega ^{16}}{2 r^{16}}
\right)\right]\, .
\end{split}
\end{equation}
Besides, it is not difficult to solve exactly the equation above using numerical methods. However, the most interesting aspect about GQTGs is that the thermodynamic properties of black holes can be determined exactly --- namely, nonperturbatively in $\gamma$ --- and analytically. 
First, expanding $f(r)$ near the horizon, according to
\begin{equation}\label{eq:nhexp}
f(r)=\frac{4\pi T}{N_k}(r-r_h)+\mathcal{O}\left((r-r_h)^2\right)\, ,
\end{equation}
and plugging this in \req{gaga}, we get two equations that relate $\omega^4$, $T$ and $r_h$:
\begin{align}
\omega^4=&k \ell^2 r_h^2+r_h^4-\frac{5 \ell^3 \gamma}{16 r_h^4}  \left(k \ell+2 x r_h\right){}^2 \left[k^2\ell^3 (3+2
   \sigma )\right.\\&\notag\left.+4 k \ell r_h
   \left(-\ell x (3+2 \sigma )+(1+\sigma ) r_h\right)+4 x r_h^2 \left(3 \ell x (1+2 \sigma )-4 (1+\sigma ) r_h\right)\right]\, ,\\ \label{omegaT}
   0=&2 r_h \left(k L^2+2 r_h \left(-\ell x+r_h\right)\right)+\frac{5 \ell^3 \gamma}{4 r_h^5}  \left(k \ell+2 x
   r_h\right){}^2 \left(k^2 \ell^3 (3+2 \sigma )\right.\\&\notag\left.+2 k \ell r_h \left(-\ell
   x \sigma +r_h+\sigma  r_h\right)-2 x (1+\sigma ) r_h^3\right)\, ,
\end{align}
where we defined again $x=\pi \ell T/N_k$. These equations are analogous to the ones in \req{TrhIIB}, but now they are exact for the theory \req{tildeS}. However, we only expect the thermodynamic relations to match in both frames at first order in $\gamma$. At that order, one finds
\begin{align}\label{TrhGQTG}
T&=\frac{N_k}{4\pi}\left[\frac{2  k}{r_h}+\frac{4 r_h}{\ell^2}+\gamma\frac{5 \left(r_h^2+k \ell^2\right){}^3 \left(k \ell^2 (\sigma +3)-2 (\sigma +1)
   r_h^2\right)}{4 \pi  \ell^2 r_h^7} \right]\, ,\\
\omega^4&=r_h^4+k\ell^2r_h^2-\frac{5\gamma}{r_h^4} \left(k L^2+r_h^2\right)^3 \left(k L^2 \sigma +r_h^2 (2
   \sigma -1)\right)\, .
\end{align}
These are different from the ones in \req{TrhIIB}. However, note that the relations $T(r_h)$  or $\omega(r_h)$ are not really physically meaningful. $T(\omega)$ is though, since $\omega^4$ is defined in both frames as (proportional to) the total energy. One can check that, at leading order in $\gamma$ this relation has the same form in both frames. 
Finally, we compute the Euclidean action, for which the same boundary terms as before are valid, namely
\begin{align}\label{S5E}
 \tilde S^{\rm E}=&-\frac{1}{16\pi G}\int_{\mathcal{M}}\diff^5 x\sqrt{|g|}\left[\frac{12}{\ell^2}+R+\frac{\zeta(3)\alpha'^3}{2\ell^2}\hat C+\frac{\zeta(3)}{8}{\alpha'}^3 \left(W^4+R^{ab}\hat{C}_{ab}\right) \right]\\
 &\notag- \frac{1}{8\pi G}\int_{\partial \mathcal{M}}\diff^{4}x\sqrt{h}\bigg[K-\frac{3}{\ell}-\frac{\ell}{4}\mathcal{R}\Bigg]\, .
\end{align}
Due to the properties of the GQTG theory, the action can be computed exactly: the Lagrangian is a total derivative and the integration only requires knowing the solution near the horizon \req{eq:nhexp} and asymptotically --- see \cite{Bueno:2018xqc} for a similar explicit computation. Since in both limits we know the exact form of $f(r)$, we obtain the following exact result
\begin{align}
\tilde S^{\rm E}=&\frac{V_k}{16 G  \ell^4 x }\left[\frac{3}{4} k^2 \ell^4+3 k \ell^2 r_h^2+r_h^3 \left(3 r_h-4 \ell x\right)\right.\\&\notag\left.-\frac{15 \ell^3 \gamma }{16 r_h^4}
   \left(k \ell+2 x r_h\right){}^3 \left(k \ell^2 (3+2 \sigma )-2 \ell x (1+2 \sigma ) r_h+4
   (1+\sigma ) r_h^2\right)\right]\, .
\end{align}
The last step is to use relation \req{omegaT}  to express $\tilde S^{\rm E}$ as a function of the temperature. We see that in general the action depends on $\sigma$. However, when we expand it at leading order in $\gamma$ the dependence on $\sigma$ disappears and we get exactly the same result as in the original frame given in  \req{SEIIB}.

\section{Discussion}\label{discu}
A summary of our findings can be found in Section \ref{summy}. We close the paper with some additional comments and conjectures. Firstly, based on the evidence presented here we state the following:

\begin{conjecture}\label{cj1} Any higher-derivative gravity Lagrangian can be mapped, order by order,
to a sum of GQTG terms by implementing redefinitions of the metric of the form \req{eq:metricredef}.
\end{conjecture}

We know there are many theories satisfying the GQTG condition \req{GQTGcond}, and the amount of terms  we can modify in the action with field redefinitions is also very large. All in all, there is so much freedom that field redefinitions seem to be able to bring the most general action \req{eq:faction} into a sum of GQTG terms, order by order in the curvature. 

Our main result is Theorem \ref{th2}, which essentially tells us that if for a given order in curvature there exists one GQTG of the form $\mathcal{L}(g_{ab},R_{abcd})$, then all densities of that type and order are completable to a GQTG. Since we know by experience that those terms exist for very high orders in curvature and general dimensions, this result virtually proves that all $\mathcal{L}($Riemann$)$ terms can be mapped to a GQTG. We would have to provide an explicit construction of these terms in order to complete a formal proof. Such systematic construction must be possible, but has not been carried out yet.

On the other hand we have seen that, interestingly, densities containing explicit covariant derivatives of the Riemann tensor do not seem to play any role. In fact, we have checked that, up to eighth order, all terms involving derivatives of the Riemann tensor are irrelevant ---they can always be mapped to other terms which already appear in the action. More generally, we have been able to prove that any term with two covariant derivatives can be completed to a GQTG which is equivalent to a GQTG of the form $\mathcal{L}(g_{ab},R_{abcd})$ when evaluated on a SSS metric. Note that the last claim is slightly different from stating that the original term can be completed to a GQTG of the form $\mathcal{L}(g_{ab},R_{abcd})$. It means that the GQTG to which the original density is completed may, in principle, contain covariant derivatives of the curvature, but it is guaranteed that those terms vanish for a SSS metric. We argued that the previous conclusion may, very likely, extend to densities with an arbitrary number of covariant derivatives, which suggests a stronger conjecture:

\begin{conjecture}\label{cj2} Any higher-derivative gravity Lagrangian can be mapped, order by order, to a sum of GQTG terms which, when evaluated on a SSS metric, are equivalent to GQTGs of the $\mathcal{L}(g_{ab},R_{abcd})$ type.
\end{conjecture}

If true, the second statement in this conjecture implies that we can study the spherically symmetric black holes  of the most general higher-derivative gravity effective action by analyzing only the solutions of the GQTGs of the form $\mathcal{L}(g_{ab},R_{abcd})$ ---like  in the example of Section \ref{typeIIb}. While, in general, the profile of the solutions will be different in every frame, recall that black hole thermodynamics is invariant under the change of frame. 
That kind of analysis was already performed in $D=4$ for a general GQTG involving arbitrarily high curvature terms \cite{PabloPablo4}. It revealed a high degree of universality for the thermodynamic behavior of the Schwarzschild black hole generalizations, including asymptotically flat stable small black holes and infinite evaporation times. Our findings here suggest that those results may actually extend to arbitrary higher-derivative theories.

The conclusion  is that theories of the GQTG class are not just toy models with interesting properties. According to our results, they capture, at the very least, a very large part of all possible effective theories of gravity, and very likely ---if Conjecture \ref{cj2} is true--- they capture all of them. From this point of view, we could think of  GQTGs  as the most general EFT expressed in a frame in which the study of spherically symmetric black holes is particularly simple and universal. 

As mentioned in Section \ref{GQTGss}, a certain subset of four-dimensional GQTGs possess second-order equations for the scale factor when evaluated on a Friedmann-Lema\^itre-Robertson-Walker ansatz, which gives rise to a well-posed cosmological evolution \cite{Arciniega:2018fxj,Cisterna:2018tgx,Arciniega:2018tnn}. The possibility that in fact all $D=4$ higher-derivative effective actions can be mapped to GQTGs belonging to this particular subset does not sound unreasonable to us and deserves further exploration. More generally, assuming Conjecture \ref{cj2} and/or Conjecture \ref{cj1} hold, one could try to impose further constraints on the GQTG family of theories targeted by the field redefinitions and then provide refinements of those conjectures.

\textbf{Note added to v2:} Recently, the existence of Quasi-topological and GQTG densities of arbitrary orders and in general dimensions was proven in \cite{Bueno:2019ycr}. This elevates the status of Conjecture \ref{cj1} to that of a Theorem for $\mathcal{L}(g_{ab},R_{abcd})$ gravities, and also for the classes of terms involving covariant derivatives of the Riemann tensor described above.

\acknowledgments
We thank Robie A. Hennigar, Alejandro Ruip\'erez, Carlos S. Shahbazi and Gonzalo Torroba for useful discussions and Stanley Deser for making us aware of reference \cite{Deser:2005pc}. The work of PB was supported by the Simons foundation through the It From Qubit Simons collaboration. The work of PAC is funded by Fundaci\'on la Caixa through a ``la Caixa - Severo Ochoa'' International pre-doctoral grant.  The work of JM is supported by the Comisi\'on Nacional de Investigaci\'on Cient\'ifica y Tecnol\'ogica (CONICYT) grant No. 21190234. The work of AM is funded by the Spanish FPU Grant No. FPU17/04964. PAC and AM were further supported by the MINECO/FEDER, UE grant PGC2018-095205-B-I00, and by the ``Centro de Excelencia Severo Ochoa" Program grant  SEV-2016-0597.

\appendix
\section{Redefining the metric}\label{App:2}

Implementing a differential change of variables directly in the action can be problematic if one is not careful enough. In order to see this, let us consider the equations of motion of $\tilde g_{ab}$ ---defined so that $g_{ab}=\tilde g_{ab}+ K_{ab}$--- by computing the variation of the new action $\tilde S[\tilde g_{ab}]=S[g_{ab}]$:\footnote{Note that in the second term we used the chain law for the functional derivative, which is in general  given by
\be
\frac{\delta S}{\delta \phi}\frac{\delta \phi}{\delta \psi}=\frac{\delta S}{\delta \phi}\frac{\partial \phi}{\partial \psi}-\partial_{a}\left(\frac{\delta S}{\delta \phi}\frac{\partial \phi}{\partial_{a}\psi}\right)+\ldots
\ee
}

\begin{equation}\label{eq:badeq}
\frac{\delta \tilde S}{\delta \tilde g_{ab}}=\frac{\delta S}{\delta g_{ab}}+\frac{\delta S}{\delta g_{ef}}\frac{\delta K_{ef}}{\delta \tilde g_{ab}} \Bigg|_{g_{ab}=\tilde g_{ab}+K_{ab}}\, .
\end{equation}
Now, it is clear that we can always solve these equations if
\begin{equation}\label{eq:goodeq}
\frac{\delta S}{\delta g_{ab}} \Bigg|_{g_{ab}=\tilde g_{ab}+K_{ab}}=0\, .
\end{equation}
In other words, implementing the change of variables directly in the equations of the original theory produces an equation that solves the equations of $\tilde S$. However, the equations of $\tilde S$ contain more solutions. These additional solutions are spurious and appear as a consequence of increasing the number of derivatives in the action, so they should not be considered. 
A possible way to formalize this intuitive argument consists in introducing auxiliary field so that the redefinition of the metric becomes algebraic. Let us consider the following action
\begin{align}
S_{\chi}=\frac{1}{16\pi G}\int \diff^D\sqrt{|g|}\Big[&-2\Lambda+R+f\left(g^{ab}, \chi_{abcd}, \chi_{e_1,abcd},\chi_{e_1e_2,abcd},\ldots\right)\\ \notag&+\frac{\partial f}{\partial \chi_{abcd}}\left(R_{abcd}-\chi_{abcd}\right)+\frac{\partial f}{\partial \chi_{e_1,abcd}}\left(\nabla_{e_1}R_{abcd}-\chi_{e_1,abcd}\right)\\ \notag &+\frac{\partial f}{\partial \chi_{e_1e_2,abcd}}\left(\nabla_{e_1}\nabla_{e_2}R_{abcd}-\chi_{e_1e_2,abcd}\right)+\ldots\Big]\, ,
\end{align}
where we have introduced some auxiliary fields $\chi_{abcd}$, $\chi_{e_{1},abcd}$, \ldots $\chi_{e_{1}\ldots e_{n},abcd}$. Let us convince ourselves that this action is equivalent to \req{eq:generalhdg}. When we take the variation with respect to $\chi_{e_{1}\ldots e_{i},abcd}$, we get
\begin{equation}
\sum_{j=0}\frac{\partial^2 f}{\partial\chi_{e_1\ldots e_i,abcd} \partial \chi_{a_1\ldots a_j,ghmn}}\left(\nabla_{a_1}\ldots\nabla_{a_j}R_{ghmn}-\chi_{a_1\ldots a_j,ghmn}\right)=0\, .
\end{equation}
In this way, we get a system of algebraic equations for the variables $\chi_{e_{1}\ldots e_{i},abcd}$ that always has the following solution
\begin{eqnarray}
\label{eq:solchi}
\chi_{abcd}&=&R_{abcd}\, ,\\
\chi_{e_1,abcd}&=&\nabla_{e_1}R_{abcd}\, ,\\
\chi_{e_1 e_2,abcd}&=&\nabla_{e_1}\nabla_{e_2}R_{abcd}\, ,\\
&\ldots&
\end{eqnarray}
This is the unique solution if the matrix of the system is invertible, and this is the expected case if $f$ is general. When we plug this solution back in the action we recover \req{eq:generalhdg} (with explicit Einstein-Hilbert and cosmological constant terms), so that both formulations are equivalent. 

Now let us perform the following redefinition of the metric in $S_{\chi}$:
\begin{equation}
g_{ab}= \tilde g_{ab}+\alpha K_{ab}\, ,\quad\text{where}\,\,\, K_{ab}=K_{ab}\left(\tilde g^{ef}, \chi_{efcd}, \chi_{a_1,efcd},\ldots\right)\, ,
\end{equation}
this is, $K_{ab}$ is a symmetric tensor formed from contractions of the $\chi$ variables and the metric, but it contains no derivatives of any field. In this way, the change of variables is algebraic and can be directly implemented in the action. We therefore get
\begin{equation}
\tilde{S}_{\chi}\left[\tilde{g}_{ab}, \chi\right]=S_{\chi}\left[\tilde{g}_{ab}+\alpha K_{ab}, \chi\right]\, ,
\end{equation}
where, for simplicity, we are collectively denoting all auxiliary variables by $\chi$.  Now, both actions are equivalent and so are the field equations:
\begin{eqnarray}
\label{eq:eqmetric}
\frac{\delta \tilde{S}_{\chi}}{\delta\tilde g_{ab}}&=&\frac{\delta S_{\chi}}{\delta g_{ab}}\bigg|_{g_{ab}= \tilde g_{ab}+\alpha K_{ab}}\, ,\\
\label{eq:eqchi}
\frac{\delta \tilde{S}_{\chi}}{\delta\chi}&=&\frac{\delta {S}_{\chi}}{\delta\chi}+\alpha \frac{\delta {S}_{\chi}}{\delta g_{ab}}\frac{\delta K_{ab}}{\delta\chi}\bigg|_{g_{ab}= \tilde g_{ab}+\alpha K_{ab}}\, .
\end{eqnarray}
Using the first equation into the second one, we see that the equations for the auxiliary variables become $\delta S_{\chi}/\delta\chi=0$, which of course have the same solution as before \req{eq:solchi}. When we take that into account, $K_{ab}$ becomes a tensor constructed from the curvature of the original metric $g_{ab}$, so that we get
\begin{equation}
g_{ab}= \tilde g_{ab}+\alpha K_{ab}\left(\tilde g^{ef}, R_{efcd}, \nabla_{\alpha_1}R_{efcd},\ldots\right)\, .
\end{equation}
Then, according to Eq.~\req{eq:eqmetric}, the equation for the metric $\tilde g_{ab}$ is simply obtained from the equation of $g_{ab}$ by substituting the change of variables:
\begin{equation}\label{eq:consistenteq}
\frac{\delta S_{\chi}}{\delta g_{ab}}\bigg|_{g_{ab}= \tilde g_{ab}+K_{ab}}=0\, .
\end{equation}
However, note that this is not the same as substituting \req{eq:solchi} in the action and taking the variation. This would yield instead
\begin{equation}\label{eq:fake}
\frac{\delta \tilde{S}_{\chi}\left[\tilde{g}_{ab}, \chi(\tilde g_{ab})\right]}{\delta\tilde g_{ab}}=\frac{\delta \tilde S_{\chi}}{\delta \tilde g_{ab}}+\frac{\delta \tilde{S}_{\chi}}{\delta\chi}\frac{\delta\chi}{\delta\tilde g_{ab}}=\frac{\delta S_{\chi}}{\delta g_{ab}}\bigg|_{g_{ab}= \tilde g_{ab}+\alpha K_{ab}}-\alpha \frac{\delta {S}_{\chi}}{\delta g_{ef}}\frac{\delta K_{ef}}{\delta\chi}\frac{\delta\chi}{\delta\tilde g_{ab}}\bigg|_{g_{ab}= \tilde g_{ab}+\alpha K_{ab}}\, .
\end{equation}
This equation is formally different to \req{eq:eqmetric} due to the second term, and it is equivalent to \req{eq:badeq}. The second term appears because the auxiliary variables $\chi(\tilde g_{\mu\nu})$ do not solve the equation $\delta \tilde{S}_{\chi}/\delta\chi=0$, but $\delta {S}_{\chi}/\delta\chi=0$. However, we must solve $\delta \tilde{S}_{\chi}/\delta\chi=0$ in order to get a solution of $\tilde{S}_{\chi}\left[\tilde{g}_{ab}, \chi\right]$, and according to \req{eq:eqchi} this would only happen if $(\delta {S}_{\chi}/\delta g_{ab}) (\partial K_{ab}/\partial\chi)=0$, so that the only consistent solutions of \req{eq:fake} are those which satisfy \req{eq:consistenteq}. 
This explains why the only solutions of \req{eq:badeq} we should consider are the ones satisfying 
\req{eq:goodeq}.

\section{$W^n \nabla W\nabla W$ terms on SSS backgrounds} \label{apricot}
In this appendix we show that \req{eq:Inv3} holds. In order to do that, it is convenient to carry out the following change or radial coordinate in the SSS ansatz \req{SSS}:
\begin{equation}
\diff \tilde{r}^2=\frac{\diff r^2}{r^2 f(r)}\, .
\end{equation}
In these coordinates, the SSS metric reads
\begin{equation}
\diff s^2=r(\tilde{r})^2\bigg [ -\tilde{N}(\tilde{r})^2\tilde{f}(\tilde{r})\diff t^2+\diff \tilde{r}^2+\diff \Omega^2_{(D-2)} \bigg],
\end{equation}
where we denoted $\tilde{N}(\tilde{r})=N(r(\tilde{r}))$ and $\tilde{f}(\tilde{r})=f(r(\tilde{r}))$.

We use a tilde to denote tensor components in the new coordinates. Direct computation shows that the components of the Weyl tensor in these new coordinates have formally the same expression as in the original ones, namely,
\begin{equation}
\tensor{ \tilde{W}}{^{ab}_{cd}}=-2\tilde{\chi}(\tilde{r}) \frac{(D-3)}{(D-1)}\tensor{ \tilde{w}}{^{ab}_{cd}}\, ,
\end{equation}
where the tensorial structure $\tensor{ \tilde{w}}{^{ab}_{cd}}$ is given by
\begin{equation}
\tensor{ \tilde{w}}{^{ab}_{cd}}=2\tilde{\tau}^{[a}_{[c} \tilde{\rho}^{b]}_{d]}-\frac{2}{(D-2)} \left(\tilde{\tau}^{[a}_{[c} \tilde{\sigma}^{b]}_{d]}+\tilde{\rho}^{[a}_{[c} \tilde{\sigma}^{b]}_{d]} \right)+\frac{2}{(D-2)(D-3)} \tilde{\sigma}^{[a}_{[c} \tilde{\sigma}^{b]}_{d]}\, ,
\end{equation}
and where $\tilde{\rho}_a^b$ denotes the projection onto our new radial coordinate $\tilde{r}$. If we define $\tilde{H}_a^b=\tilde{\tau}_a^b+\tilde{\rho}_a^b$, we may express $\tensor{ \tilde{w}}{^{ab}_{cd}}$ as
\begin{equation}
\tensor{ \tilde{w}}{^{ab}_{cd}}=\tilde{H}^{[a}_{[c} \tilde{H}^{b]}_{d]}-\frac{2}{(D-2)} \tilde{H}^{[a}_{[c} \tilde{\sigma}^{b]}_{d]}+\frac{2}{(D-2)(D-3)} \tilde{\sigma}^{[a}_{[c} \tilde{\sigma}^{b]}_{d]}\, .
\label{mywayw}
\end{equation}
Consequently, the covariant derivative of the Weyl tensor turns out to be
\begin{equation}
\left. \nabla_e \tensor{\tilde{W}}{^{ab}_{cd}}  \right|_{\rm SSS}=-2\frac{(D-3)}{(D-1)} \left[ \frac{\diff \tilde{\chi}}{\diff \tilde{r}} \delta_e^1 \tensor{ \tilde{w}}{^{ab}_{cd}}+\tilde{\chi}(\tilde{r}) \left. {\nabla}_e \tensor{ \tilde{w}}{^{ab}_{cd}} \right|_{\rm SSS} \right]\, ,
\end{equation}
where we are denoting the components of the covariant derivative of any tensor $T$ in our new coordinates as $\nabla_e \tilde{T}_{ab...}^{cd...}$. Hence we just need to work out $\left. {\nabla}_e \tensor{ \tilde{w}}{^{ab}_{cd}} \right|_{\rm SSS}$. Using \eqref{mywayw}, we find
\begin{equation}
\begin{split}
\nabla_e \tensor{ \tilde{w}}{^{ab}_{cd}}&=2 \nabla_e \tilde{H}^{[a}_{[c} \tilde{H}^{b]}_{d]}-\frac{2}{(D-2)} \nabla_e \tilde{H}^{[a}_{[c} \tilde{\sigma}^{b]}_{d]}\\&-\frac{2}{(D-2)} \nabla_e \tilde{\sigma}^{[a}_{[c} \tilde{H}^{b]}_{d]}+\frac{4}{(D-2)(D-3)} \nabla_e \tilde{\sigma}^{[a}_{[c} \tilde{\sigma}^{b]}_{d]}\, .
\end{split}
\end{equation}
Since $\nabla_e \tilde{H}_a^b+\nabla_e \tilde{\sigma}_a^b=0$, we just need to compute $\nabla_e \tilde{H}_a^b$. A straightforward calculation produces
\begin{equation}
\nabla_e \tilde{H}_a^b=\frac{1}{(r(\tilde{r}))^3}\frac{\diff r}{\diff \tilde{r}} \tilde{g}_{eg} \tilde{\sigma}_a^f \delta_1^b +\frac{1}{r(\tilde{r})}\frac{\diff r}{\diff \tilde{r}} (D-2) \tilde{\sigma}_e^b \delta_a^1\, .
\end{equation}
Using this, the covariant derivative of the Weyl tensor gives
\begin{align}
\label{cdwtab}
\left. \nabla_e \tensor{ \tilde{W}}{^{ab}_{cd}}  \right|_{\rm SSS}=&-2\frac{(D-3)}{(D-1)}  \frac{\diff \tilde{\chi}}{\diff\tilde{r}} \delta_e^1 \tensor{ \tilde{w}}{^{ab}_{cd}} -2\frac{(D-3)}{(D-1)} \tilde{\chi}(\tilde{r}) \frac{\diff r}{\diff\tilde{r}} \bigg[ \frac{2}{(r(\tilde{r}))^3} \tilde{g}_{ef} \tilde{\sigma}_{[c|}^{f} \delta_1^{[a} \tilde{H}_{| d]}^{b]}\\ \notag &+\frac{2(D-2)}{r(\tilde{r})} \tilde{\sigma}_e^{[a|} \delta_{[c}^1 \tilde{H}_{d]}^{|b]}-\frac{2 }{(D-2)(r(\tilde{r}))^3}\tilde{g}_{ef} \tilde{\sigma}_{[c|}^{f} \delta_1^{[a} \tilde{\sigma}_{| d]}^{b]}\\ \notag &-\frac{2}{r(\tilde{r})}\tilde{\sigma}_e^{[a|} \delta_{[c}^1 \tilde{\sigma}_{d]}^{|b]}  +\frac{2 }{(D-2)(r(\tilde{r}))^3}\tilde{g}_{ef} \tilde{\sigma}_{[c|}^{f} \delta_1^{[a} \tilde{H}_{| d]}^{b]}+\frac{2 }{r(\tilde{r})}\tilde{\sigma}_e^{[a|} \delta_{[c}^1 \tilde{H}_{d]}^{|b]}\\  \notag&-\frac{4}{(D-2)(D-3)(r(\tilde{r}))^3}\tilde{g}_{ef} \tilde{\sigma}_{[c|}^{f} \delta_1^{[a} \tilde{\sigma}_{| d]}^{b]}-\frac{4 }{(D-3)r(\tilde{r})}\tilde{\sigma}_e^{[a|} \delta_{[c}^1 \tilde{\sigma}_{d]}^{|b]} \bigg]\, .
\end{align}
Equipped with \eqref{cdwtab}, we may infer the general form of any invariant $\left. \mathcal{R}_2^{\{1,1\}} \right |_{\rm SSS}$ as defined in \req{eq:Inv2}. Since the $\left. \mathcal{R}_2^{\{1,1\}} \right |_{\rm SSS}$ are scalars, we can obtain them expressed in the original coordinates by performing all calculations in the new ones and then substituting any dependence on $\tilde{r}$ by the initial radial coordinate $r$.

We notice the following facts:
a) any $\left. \mathcal{R}_2^{\{1,1\}} \right |_{\rm SSS}$ will have three types of terms: those carrying  a factor $\tilde{\chi}^n \left (\diff \tilde{\chi}/\diff \tilde{r} \right)^2$,  those involving a factor $\tilde{\chi}^{n+1}(\diff \tilde{\chi}/\diff \tilde{r}) (\diff r/\diff\tilde{r})$ and a third type of terms with the common factor $\tilde{\chi}^{n+2}(\diff r/\diff \tilde{r}  )^2$; b) since $\tilde{r}$ is dimensionless, we infer that the first type of terms is not weighted by any power of $r$, the second type is accompanied by $r^{-1}$ and the third type, by $r^{-2}$. An additional overall factor of $r^{-2}$ is required by dimensional analysis.
Using these observations, it follows that
\begin{equation}
\left. \mathcal{R}_2^{\{1,1\}} \right |_{\rm SSS}=\frac{\tilde{\chi}^n(\tilde{r})}{r(\tilde{r})^2}\bigg[ c_1 \left (\frac{\diff \tilde{\chi}}{\diff \tilde{r}} \right)^2+c_2  \frac{\diff \tilde{\chi}}{\diff\tilde{r}} \frac{\diff r}{\diff\tilde{r}} \frac{\tilde{\chi}(\tilde{r})}{r(\tilde{r})}+c_3  \left ( \frac{\tilde{\chi}(\tilde{r})}{r(\tilde{r})} \right )^2 \left( \frac{\diff r}{\diff \tilde{r}}  \right)^2\bigg]\, ,
\end{equation}
for some constants $c_1$, $c_2$, $c_3$ which will depend on the specific term.
Taking into account that $\diff \tilde{r}/\diff r =1/ (r \sqrt{f(r)})$
we finally find
\begin{equation}
\left. \mathcal{R}_2^{\{1,1\}} \right |_{\rm SSS}=\chi^n f(r)\left ( c_1  (\chi ')^2+c_2\frac{\chi \chi'}{r}+c_3 \frac{\chi^2}{r^2} \right ),
\end{equation}
where $\chi'=\diff \chi/\diff r$.

\bibliography{Gravities}

\newcommand{\noop}[1]{}
\providecommand{\href}[2]{#2}\begingroup\raggedright\begin{thebibliography}{100}

\bibitem{Gross:1986mw}
D.~J. Gross and J.~H. Sloan, {\it {The Quartic Effective Action for the
  Heterotic String}},  {\em Nucl. Phys.} {\bf B291} (1987) 41--89.

\bibitem{Gross:1986iv}
D.~J. Gross and E.~Witten, {\it {Superstring Modifications of Einstein's
  Equations}},  {\em Nucl. Phys.} {\bf B277} (1986) 1.

\bibitem{Grisaru:1986vi}
M.~T. Grisaru and D.~Zanon, {\it {$\sigma$ Model Superstring Corrections to the
  Einstein-hilbert Action}},  {\em Phys. Lett.} {\bf B177} (1986) 347--351.

\bibitem{Grisaru:1986px}
M.~T. Grisaru, A.~E.~M. van~de Ven and D.~Zanon, {\it {Four Loop beta Function
  for the N=1 and N=2 Supersymmetric Nonlinear Sigma Model in Two-Dimensions}},
   {\em Phys. Lett.} {\bf B173} (1986) 423--428.

\bibitem{Bergshoeff:1989de}
E.~A. Bergshoeff and M.~de~Roo, {\it {The Quartic Effective Action of the
  Heterotic String and Supersymmetry}},  {\em Nucl. Phys.} {\bf B328} (1989)
  439--468.

\bibitem{Green:2003an}
M.~B. Green and C.~Stahn, {\it {D3-branes on the Coulomb branch and
  instantons}},  {\em JHEP} {\bf 09} (2003) 052
  [\href{http://arXiv.org/abs/hep-th/0308061}{{\tt hep-th/0308061}}].

\bibitem{Frolov:2001xr}
S.~Frolov, I.~R. Klebanov and A.~A. Tseytlin, {\it {String corrections to the
  holographic RG flow of supersymmetric SU(N) x SU(N + M) gauge theory}},  {\em
  Nucl. Phys.} {\bf B620} (2002) 84--108
  [\href{http://arXiv.org/abs/hep-th/0108106}{{\tt hep-th/0108106}}].

\bibitem{Birrell:1982ix}
N.~D. Birrell and P.~C.~W. Davies, {\em {Quantum Fields in Curved Space}}.
\newblock Cambridge Monographs on Mathematical Physics. Cambridge Univ. Press,
  Cambridge, UK, 1984.

\bibitem{DeWitt:1967yk}
B.~S. DeWitt, {\it {Quantum Theory of Gravity. 1. The Canonical Theory}},  {\em
  Phys. Rev.} {\bf 160} (1967) 1113--1148. [3,93(1987)].

\bibitem{tHooft:1974toh}
G.~'t~Hooft and M.~J.~G. Veltman, {\it {One loop divergencies in the theory of
  gravitation}},  {\em Ann. Inst. H. Poincare Phys. Theor.} {\bf A20} (1974)
  69--94.

\bibitem{Goroff:1985th}
M.~H. Goroff and A.~Sagnotti, {\it {The Ultraviolet Behavior of Einstein
  Gravity}},  {\em Nucl. Phys.} {\bf B266} (1986) 709--736.

\bibitem{vandeVen:1991gw}
A.~E.~M. van~de Ven, {\it {Two loop quantum gravity}},  {\em Nucl. Phys.} {\bf
  B378} (1992) 309--366.

\bibitem{Solodukhin:2015ypa}
S.~N. Solodukhin, {\it {Metric Redefinition and UV Divergences in Quantum
  Einstein Gravity}},  {\em Phys. Lett.} {\bf B754} (2016) 157--161
  [\href{http://arXiv.org/abs/1509.04890}{{\tt 1509.04890}}].

\bibitem{Weinberg:1995mt}
S.~Weinberg, {\em {The Quantum theory of fields. Vol. 1: Foundations}}.
\newblock Cambridge University Press, 2005.

\bibitem{Lovelock1}
D.~Lovelock, {\it Divergence-free tensorial concomitants},  {\em aequationes
  mathematicae} {\bf 4} (1970), no.~1 127--138.

\bibitem{Lovelock:1971yv}
D.~Lovelock, {\it {The Einstein tensor and its generalizations}},  {\em J.
  Math. Phys.} {\bf 12} (1971) 498--501.

\bibitem{Stelle:1977ry}
K.~S. Stelle, {\it {Classical Gravity with Higher Derivatives}},  {\em Gen.
  Rel. Grav.} {\bf 9} (1978) 353--371.

\bibitem{Stelle:1976gc}
K.~S. Stelle, {\it {Renormalization of Higher Derivative Quantum Gravity}},
  {\em Phys. Rev.} {\bf D16} (1977) 953--969.

\bibitem{Lu}
H.~Lu and C.~N. Pope, {\it {Critical Gravity in Four Dimensions}},  {\em Phys.
  Rev. Lett.} {\bf 106} (2011) 181302
  [\href{http://arXiv.org/abs/1101.1971}{{\tt 1101.1971}}].

\bibitem{Tekin3}
A.~Karasu, E.~Kenar and B.~Tekin, {\it {Minimal extension of Einstein‚Äôs
  theory: The quartic gravity}},  {\em Phys. Rev.} {\bf D93} (2016), no.~8
  084040 [\href{http://arXiv.org/abs/1602.02567}{{\tt 1602.02567}}].

\bibitem{Maldacena:2011mk}
J.~Maldacena, {\it {Einstein Gravity from Conformal Gravity}},
  \href{http://arXiv.org/abs/1105.5632}{{\tt 1105.5632}}.

\bibitem{Oliva:2011xu}
J.~Oliva and S.~Ray, {\it {Birkhoff's Theorem in Higher Derivative Theories of
  Gravity}},  {\em Class. Quant. Grav.} {\bf 28} (2011) 175007
  [\href{http://arXiv.org/abs/1104.1205}{{\tt 1104.1205}}].

\bibitem{Oliva:2012zs}
J.~Oliva and S.~Ray, {\it {Birkhoff's Theorem in Higher Derivative Theories of
  Gravity II}},  {\em Phys. Rev.} {\bf D86} (2012) 084014
  [\href{http://arXiv.org/abs/1201.5601}{{\tt 1201.5601}}].

\bibitem{Quasi2}
J.~Oliva and S.~Ray, {\it {A new cubic theory of gravity in five dimensions:
  Black hole, Birkhoff's theorem and C-function}},  {\em Class. Quant. Grav.}
  {\bf 27} (2010) 225002 [\href{http://arXiv.org/abs/1003.4773}{{\tt
  1003.4773}}].

\bibitem{PabloPablo}
P.~Bueno and P.~A. Cano, {\it {Einsteinian cubic gravity}},  {\em Phys. Rev.}
  {\bf D94} (2016), no.~10 104005 [\href{http://arXiv.org/abs/1607.06463}{{\tt
  1607.06463}}].

\bibitem{Lu:2015cqa}
H.~Lu, A.~Perkins, C.~N. Pope and K.~S. Stelle, {\it {Black Holes in
  Higher-Derivative Gravity}},  {\em Phys. Rev. Lett.} {\bf 114} (2015), no.~17
  171601 [\href{http://arXiv.org/abs/1502.01028}{{\tt 1502.01028}}].

\bibitem{Tekin2}
T.~C. Sisman, I.~Gullu and B.~Tekin, {\it {All unitary cubic curvature
  gravities in D dimensions}},  {\em Class. Quant. Grav.} {\bf 28} (2011)
  195004 [\href{http://arXiv.org/abs/1103.2307}{{\tt 1103.2307}}].

\bibitem{Aspects}
P.~Bueno, P.~A. Cano, V.~S. Min and M.~R. Visser, {\it {Aspects of general
  higher-order gravities}},  {\em Phys. Rev.} {\bf D95} (2017), no.~4 044010
  [\href{http://arXiv.org/abs/1610.08519}{{\tt 1610.08519}}].

\bibitem{Anastasiou:2016jix}
G.~Anastasiou and R.~Olea, {\it {From conformal to Einstein Gravity}},  {\em
  Phys. Rev.} {\bf D94} (2016), no.~8 086008
  [\href{http://arXiv.org/abs/1608.07826}{{\tt 1608.07826}}].

\bibitem{Sotiriou}
T.~P. Sotiriou and V.~Faraoni, {\it {f(R) Theories Of Gravity}},  {\em Rev.
  Mod. Phys.} {\bf 82} (2010) 451--497
  [\href{http://arXiv.org/abs/0805.1726}{{\tt 0805.1726}}].

\bibitem{Maldacena}
J.~M. Maldacena, {\it {The Large N limit of superconformal field theories and
  supergravity}},  {\em Int. J. Theor. Phys.} {\bf 38} (1999) 1113--1133
  [\href{http://arXiv.org/abs/hep-th/9711200}{{\tt hep-th/9711200}}]. [Adv.
  Theor. Math. Phys.2,231(1998)].

\bibitem{Witten}
E.~Witten, {\it {Anti-de Sitter space and holography}},  {\em Adv. Theor. Math.
  Phys.} {\bf 2} (1998) 253--291
  [\href{http://arXiv.org/abs/hep-th/9802150}{{\tt hep-th/9802150}}].

\bibitem{Hofman:2008ar}
D.~M. Hofman and J.~Maldacena, {\it {Conformal collider physics: Energy and
  charge correlations}},  {\em JHEP} {\bf 05} (2008) 012
  [\href{http://arXiv.org/abs/0803.1467}{{\tt 0803.1467}}].

\bibitem{Camanho:2009vw}
X.~O. Camanho and J.~D. Edelstein, {\it {Causality constraints in AdS/CFT from
  conformal collider physics and Gauss-Bonnet gravity}},  {\em JHEP} {\bf 04}
  (2010) 007 [\href{http://arXiv.org/abs/0911.3160}{{\tt 0911.3160}}].

\bibitem{deBoer:2009pn}
J.~de~Boer, M.~Kulaxizi and A.~Parnachev, {\it {AdS(7)/CFT(6), Gauss-Bonnet
  Gravity, and Viscosity Bound}},  {\em JHEP} {\bf 03} (2010) 087
  [\href{http://arXiv.org/abs/0910.5347}{{\tt 0910.5347}}].

\bibitem{Buchel:2009sk}
A.~Buchel, J.~Escobedo, R.~C. Myers, M.~F. Paulos, A.~Sinha and M.~Smolkin,
  {\it {Holographic GB gravity in arbitrary dimensions}},  {\em JHEP} {\bf 03}
  (2010) 111 [\href{http://arXiv.org/abs/0911.4257}{{\tt 0911.4257}}].

\bibitem{deBoer:2009gx}
J.~de~Boer, M.~Kulaxizi and A.~Parnachev, {\it {Holographic Lovelock Gravities
  and Black Holes}},  {\em JHEP} {\bf 06} (2010) 008
  [\href{http://arXiv.org/abs/0912.1877}{{\tt 0912.1877}}].

\bibitem{Grozdanov:2014kva}
S.~Grozdanov and A.~O. Starinets, {\it {On the universal identity in second
  order hydrodynamics}},  {\em JHEP} {\bf 03} (2015) 007
  [\href{http://arXiv.org/abs/1412.5685}{{\tt 1412.5685}}].

\bibitem{Andrade:2016rln}
T.~Andrade, J.~Casalderrey-Solana and A.~Ficnar, {\it {Holographic
  Isotropisation in Gauss-Bonnet Gravity}},  {\em JHEP} {\bf 02} (2017) 016
  [\href{http://arXiv.org/abs/1610.08987}{{\tt 1610.08987}}].

\bibitem{Konoplya:2017zwo}
R.~A. Konoplya and A.~Zhidenko, {\it {Quasinormal modes of Gauss-Bonnet-AdS
  black holes: towards holographic description of finite coupling}},  {\em
  JHEP} {\bf 09} (2017) 139 [\href{http://arXiv.org/abs/1705.07732}{{\tt
  1705.07732}}].

\bibitem{Quasi}
R.~C. Myers and B.~Robinson, {\it {Black Holes in Quasi-topological Gravity}},
  {\em JHEP} {\bf 08} (2010) 067 [\href{http://arXiv.org/abs/1003.5357}{{\tt
  1003.5357}}].

\bibitem{Myers:2010jv}
R.~C. Myers, M.~F. Paulos and A.~Sinha, {\it {Holographic studies of
  quasi-topological gravity}},  {\em JHEP} {\bf 08} (2010) 035
  [\href{http://arXiv.org/abs/1004.2055}{{\tt 1004.2055}}].

\bibitem{HoloRen}
L.-Y. Hung, R.~C. Myers, M.~Smolkin and A.~Yale, {\it {Holographic Calculations
  of Renyi Entropy}},  {\em JHEP} {\bf 12} (2011) 047
  [\href{http://arXiv.org/abs/1110.1084}{{\tt 1110.1084}}].

\bibitem{Hung:2014npa}
L.-Y. Hung, R.~C. Myers and M.~Smolkin, {\it {Twist operators in higher
  dimensions}},  {\em JHEP} {\bf 10} (2014) 178
  [\href{http://arXiv.org/abs/1407.6429}{{\tt 1407.6429}}].

\bibitem{deBoer:2011wk}
J.~de~Boer, M.~Kulaxizi and A.~Parnachev, {\it {Holographic Entanglement
  Entropy in Lovelock Gravities}},  {\em JHEP} {\bf 07} (2011) 109
  [\href{http://arXiv.org/abs/1101.5781}{{\tt 1101.5781}}].

\bibitem{Bianchi:2016xvf}
L.~Bianchi, S.~Chapman, X.~Dong, D.~A. Galante, M.~Meineri and R.~C. Myers,
  {\it {Shape dependence of holographic Rényi entropy in general dimensions}},
   {\em JHEP} {\bf 11} (2016) 180 [\href{http://arXiv.org/abs/1607.07418}{{\tt
  1607.07418}}].

\bibitem{Bueno:2018xqc}
P.~Bueno, P.~A. Cano and A.~Ruipérez, {\it {Holographic studies of Einsteinian
  cubic gravity}},  {\em JHEP} {\bf 03} (2018) 150
  [\href{http://arXiv.org/abs/1802.00018}{{\tt 1802.00018}}].

\bibitem{Myers:2010tj}
R.~C. Myers and A.~Sinha, {\it {Holographic c-theorems in arbitrary
  dimensions}},  {\em JHEP} {\bf 01} (2011) 125
  [\href{http://arXiv.org/abs/1011.5819}{{\tt 1011.5819}}].

\bibitem{Myers:2010xs}
R.~C. Myers and A.~Sinha, {\it {Seeing a c-theorem with holography}},  {\em
  Phys. Rev.} {\bf D82} (2010) 046006
  [\href{http://arXiv.org/abs/1006.1263}{{\tt 1006.1263}}].

\bibitem{Mezei:2014zla}
M.~Mezei, {\it {Entanglement entropy across a deformed sphere}},  {\em Phys.
  Rev.} {\bf D91} (2015), no.~4 045038
  [\href{http://arXiv.org/abs/1411.7011}{{\tt 1411.7011}}].

\bibitem{Bueno1}
P.~Bueno, R.~C. Myers and W.~Witczak-Krempa, {\it {Universality of corner
  entanglement in conformal field theories}},  {\em Phys. Rev. Lett.} {\bf 115}
  (2015) 021602 [\href{http://arXiv.org/abs/1505.04804}{{\tt 1505.04804}}].

\bibitem{Bueno2}
P.~Bueno and R.~C. Myers, {\it {Corner contributions to holographic
  entanglement entropy}},  {\em JHEP} {\bf 08} (2015) 068
  [\href{http://arXiv.org/abs/1505.07842}{{\tt 1505.07842}}].

\bibitem{Buchel:2004di}
A.~Buchel, J.~T. Liu and A.~O. Starinets, {\it {Coupling constant dependence of
  the shear viscosity in N=4 supersymmetric Yang-Mills theory}},  {\em Nucl.
  Phys.} {\bf B707} (2005) 56--68
  [\href{http://arXiv.org/abs/hep-th/0406264}{{\tt hep-th/0406264}}].

\bibitem{Kats:2007mq}
Y.~Kats and P.~Petrov, {\it {Effect of curvature squared corrections in AdS on
  the viscosity of the dual gauge theory}},  {\em JHEP} {\bf 01} (2009) 044
  [\href{http://arXiv.org/abs/0712.0743}{{\tt 0712.0743}}].

\bibitem{Brigante:2007nu}
M.~Brigante, H.~Liu, R.~C. Myers, S.~Shenker and S.~Yaida, {\it {Viscosity
  Bound Violation in Higher Derivative Gravity}},  {\em Phys. Rev.} {\bf D77}
  (2008) 126006 [\href{http://arXiv.org/abs/0712.0805}{{\tt 0712.0805}}].

\bibitem{Myers:2008yi}
R.~C. Myers, M.~F. Paulos and A.~Sinha, {\it {Quantum corrections to eta/s}},
  {\em Phys. Rev.} {\bf D79} (2009) 041901
  [\href{http://arXiv.org/abs/0806.2156}{{\tt 0806.2156}}].

\bibitem{Cai:2008ph}
R.-G. Cai, Z.-Y. Nie and Y.-W. Sun, {\it {Shear Viscosity from Effective
  Couplings of Gravitons}},  {\em Phys. Rev.} {\bf D78} (2008) 126007
  [\href{http://arXiv.org/abs/0811.1665}{{\tt 0811.1665}}].

\bibitem{Ge:2008ni}
X.-H. Ge, Y.~Matsuo, F.-W. Shu, S.-J. Sin and T.~Tsukioka, {\it {Viscosity
  Bound, Causality Violation and Instability with Stringy Correction and
  Charge}},  {\em JHEP} {\bf 10} (2008) 009
  [\href{http://arXiv.org/abs/0808.2354}{{\tt 0808.2354}}].

\bibitem{Jacobson:1993vj}
T.~Jacobson, G.~Kang and R.~C. Myers, {\it {On black hole entropy}},  {\em
  Phys. Rev.} {\bf D49} (1994) 6587--6598
  [\href{http://arXiv.org/abs/gr-qc/9312023}{{\tt gr-qc/9312023}}].

\bibitem{Hennigar:2016gkm}
R.~A. Hennigar and R.~B. Mann, {\it {Black holes in Einsteinian cubic
  gravity}},  {\em Phys. Rev.} {\bf D95} (2017), no.~6 064055
  [\href{http://arXiv.org/abs/1610.06675}{{\tt 1610.06675}}].

\bibitem{PabloPablo2}
P.~Bueno and P.~A. Cano, {\it {Four-dimensional black holes in Einsteinian
  cubic gravity}},  {\em Phys. Rev.} {\bf D94} (2016), no.~12 124051
  [\href{http://arXiv.org/abs/1610.08019}{{\tt 1610.08019}}].

\bibitem{Hennigar:2017ego}
R.~A. Hennigar, D.~Kubizňák and R.~B. Mann, {\it {Generalized
  quasitopological gravity}},  {\em Phys. Rev.} {\bf D95} (2017), no.~10 104042
  [\href{http://arXiv.org/abs/1703.01631}{{\tt 1703.01631}}].

\bibitem{PabloPablo3}
P.~Bueno and P.~A. Cano, {\it {On black holes in higher-derivative gravities}},
   {\em Class. Quant. Grav.} {\bf 34} (2017), no.~17 175008
  [\href{http://arXiv.org/abs/1703.04625}{{\tt 1703.04625}}].

\bibitem{Ahmed:2017jod}
J.~Ahmed, R.~A. Hennigar, R.~B. Mann and M.~Mir, {\it {Quintessential Quartic
  Quasi-topological Quartet}},  {\em JHEP} {\bf 05} (2017) 134
  [\href{http://arXiv.org/abs/1703.11007}{{\tt 1703.11007}}].

\bibitem{PabloPablo4}
P.~Bueno and P.~A. Cano, {\it {Universal black hole stability in four
  dimensions}},  {\em Phys. Rev.} {\bf D96} (2017), no.~2 024034
  [\href{http://arXiv.org/abs/1704.02967}{{\tt 1704.02967}}].

\bibitem{Tekin1}
B.~Tekin, {\it {Particle Content of Quadratic and $f(R_{\mu\nu\sigma \rho})$
  Theories in $(A)dS$}},  {\em Phys. Rev.} {\bf D93} (2016), no.~10 101502
  [\href{http://arXiv.org/abs/1604.00891}{{\tt 1604.00891}}].

\bibitem{Love}
P.~Bueno, P.~A. Cano, A.~O. Lasso and P.~F. Ramirez, {\it {f(Lovelock) theories
  of gravity}},  {\em JHEP} {\bf 04} (2016) 028
  [\href{http://arXiv.org/abs/1602.07310}{{\tt 1602.07310}}].

\bibitem{Ghodsi:2017iee}
A.~Ghodsi and F.~Najafi, {\it {Ricci cubic gravity in d dimensions, gravitons
  and SAdS/Lifshitz black holes}},  {\em Eur. Phys. J.} {\bf C77} (2017), no.~8
  559 [\href{http://arXiv.org/abs/1702.06798}{{\tt 1702.06798}}].

\bibitem{Li:2017ncu}
Y.-Z. Li, H.-S. Liu and H.~Lu, {\it {Quasi-Topological Ricci Polynomial
  Gravities}},  \href{http://arXiv.org/abs/1708.07198}{{\tt 1708.07198}}.

\bibitem{Li:2017txk}
Y.-Z. Li, H.~Lu and J.-B. Wu, {\it {Causality and a-theorem Constraints on
  Ricci Polynomial and Riemann Cubic Gravities}},  {\em Phys. Rev.} {\bf D97}
  (2018), no.~2 024023 [\href{http://arXiv.org/abs/1711.03650}{{\tt
  1711.03650}}].

\bibitem{Li:2018drw}
Y.-Z. Li, H.~Lü and Z.-F. Mai, {\it {Universal Structure of Covariant
  Holographic Two-Point Functions In Massless Higher-Order Gravities}},  {\em
  JHEP} {\bf 10} (2018) 063 [\href{http://arXiv.org/abs/1808.00494}{{\tt
  1808.00494}}].

\bibitem{Li:2019auk}
Y.-Z. Li, {\it {Holographic Studies of The Generic Massless Cubic Gravities}},
  {\em Phys. Rev.} {\bf D99} (2019), no.~6 066014
  [\href{http://arXiv.org/abs/1901.03349}{{\tt 1901.03349}}].

\bibitem{Lu:2019urr}
H.~Lu and R.~Wen, {\it {Holographic (a,c)-charges and Their Universal Relation
  in d=6 from Massless Higher-order Gravities}},
  \href{http://arXiv.org/abs/1901.11037}{{\tt 1901.11037}}.

\bibitem{Dehghani:2011vu}
M.~H. Dehghani, A.~Bazrafshan, R.~B. Mann, M.~R. Mehdizadeh, M.~Ghanaatian and
  M.~H. Vahidinia, {\it {Black Holes in Quartic Quasitopological Gravity}},
  {\em Phys. Rev.} {\bf D85} (2012) 104009
  [\href{http://arXiv.org/abs/1109.4708}{{\tt 1109.4708}}].

\bibitem{Cisterna:2017umf}
A.~Cisterna, L.~Guajardo, M.~Hassaine and J.~Oliva, {\it {Quintic
  quasi-topological gravity}},  {\em JHEP} {\bf 04} (2017) 066
  [\href{http://arXiv.org/abs/1702.04676}{{\tt 1702.04676}}].

\bibitem{Arnowitt:1960es}
R.~L. Arnowitt, S.~Deser and C.~W. Misner, {\it {Canonical variables for
  general relativity}},  {\em Phys. Rev.} {\bf 117} (1960) 1595--1602.

\bibitem{Arnowitt:1960zzc}
R.~Arnowitt, S.~Deser and C.~W. Misner, {\it {Energy and the Criteria for
  Radiation in General Relativity}},  {\em Phys. Rev.} {\bf 118} (1960)
  1100--1104.

\bibitem{Arnowitt:1961zz}
R.~L. Arnowitt, S.~Deser and C.~W. Misner, {\it {Coordinate invariance and
  energy expressions in general relativity}},  {\em Phys. Rev.} {\bf 122}
  (1961) 997.

\bibitem{Bueno:2018uoy}
P.~Bueno, P.~A. Cano, R.~A. Hennigar and R.~B. Mann, {\it {NUTs and bolts
  beyond Lovelock}},  {\em JHEP} {\bf 10} (2018) 095
  [\href{http://arXiv.org/abs/1808.01671}{{\tt 1808.01671}}].

\bibitem{Arciniega:2018fxj}
G.~Arciniega, J.~D. Edelstein and L.~G. Jaime, {\it {Towards purely geometric
  inflation and late time acceleration}},
  \href{http://arXiv.org/abs/1810.08166}{{\tt 1810.08166}}.

\bibitem{Cisterna:2018tgx}
A.~Cisterna, N.~Grandi and J.~Oliva, {\it {On four-dimensional Einsteinian
  gravity, quasitopological gravity, cosmology and black holes}},
  \href{http://arXiv.org/abs/1811.06523}{{\tt 1811.06523}}.

\bibitem{Arciniega:2018tnn}
G.~Arciniega, P.~Bueno, P.~A. Cano, J.~D. Edelstein, R.~A. Hennigar and L.~G.
  Jaime, {\it {Geometric Inflation}},
  \href{http://arXiv.org/abs/1812.11187}{{\tt 1812.11187}}.

\bibitem{Dey:2016pei}
A.~Dey, P.~Roy and T.~Sarkar, {\it {On Holographic R\'enyi Entropy in Some
  Modified Theories of Gravity}},  \href{http://arXiv.org/abs/1609.02290}{{\tt
  1609.02290}}.

\bibitem{Feng:2017tev}
X.-H. Feng, H.~Huang, Z.-F. Mai and H.~Lu, {\it {Bounce Universe and Black
  Holes from Critical Einsteinian Cubic Gravity}},  {\em Phys. Rev.} {\bf D96}
  (2017), no.~10 104034 [\href{http://arXiv.org/abs/1707.06308}{{\tt
  1707.06308}}].

\bibitem{Hennigar:2017umz}
R.~A. Hennigar, {\it {Criticality for charged black branes}},  {\em JHEP} {\bf
  09} (2017) 082 [\href{http://arXiv.org/abs/1705.07094}{{\tt 1705.07094}}].

\bibitem{Hennigar:2018hza}
R.~A. Hennigar, M.~B.~J. Poshteh and R.~B. Mann, {\it {Shadows, Signals, and
  Stability in Einsteinian Cubic Gravity}},  {\em Phys. Rev.} {\bf D97} (2018),
  no.~6 064041 [\href{http://arXiv.org/abs/1801.03223}{{\tt 1801.03223}}].

\bibitem{Bueno:2018yzo}
P.~Bueno, P.~A. Cano, R.~A. Hennigar and R.~B. Mann, {\it {Universality of
  Squashed-Sphere Partition Functions}},  {\em Phys. Rev. Lett.} {\bf 122}
  (2019), no.~7 071602 [\href{http://arXiv.org/abs/1808.02052}{{\tt
  1808.02052}}].

\bibitem{Poshteh:2018wqy}
M.~B.~J. Poshteh and R.~B. Mann, {\it {Gravitational Lensing by Black Holes in
  Einsteinian Cubic Gravity}},  {\em Phys. Rev.} {\bf D99} (2019), no.~2 024035
  [\href{http://arXiv.org/abs/1810.10657}{{\tt 1810.10657}}].

\bibitem{Mir:2019ecg}
M.~Mir, R.~A. Hennigar, J.~Ahmed and R.~B. Mann, {\it {Black hole chemistry and
  holography in generalized quasi-topological gravity}},
  \href{http://arXiv.org/abs/1902.02005}{{\tt 1902.02005}}.

\bibitem{Mir:2019rik}
M.~Mir and R.~B. Mann, {\it {On generalized quasi-topological cubic-quartic
  gravity: thermodynamics and holography}},
  \href{http://arXiv.org/abs/1902.10906}{{\tt 1902.10906}}.

\bibitem{Mehdizadeh:2019qvc}
M.~R. Mehdizadeh and A.~H. Ziaie, {\it {Traversable wormholes in Einsteinian
  cubic gravity}},  \href{http://arXiv.org/abs/1903.10907}{{\tt 1903.10907}}.

\bibitem{Erices:2019mkd}
C.~Erices, E.~Papantonopoulos and E.~N. Saridakis, {\it {Cosmology in cubic and
  $f(P)$ gravity}},  \href{http://arXiv.org/abs/1903.11128}{{\tt 1903.11128}}.

\bibitem{Emond:2019crr}
W.~T. Emond and N.~Moynihan, {\it {Scattering Amplitudes, Black Holes and
  Leading Singularities in Cubic Theories of Gravity}},
  \href{http://arXiv.org/abs/1905.08213}{{\tt 1905.08213}}.

\bibitem{Gibbons:1976ue}
G.~W. Gibbons and S.~W. Hawking, {\it {Action Integrals and Partition Functions
  in Quantum Gravity}},  {\em Phys. Rev.} {\bf D15} (1977) 2752--2756.

\bibitem{Wald:1993nt}
R.~M. Wald, {\it {Black hole entropy is the Noether charge}},  {\em Phys. Rev.}
  {\bf D48} (1993) 3427--3431 [\href{http://arXiv.org/abs/gr-qc/9307038}{{\tt
  gr-qc/9307038}}].

\bibitem{Exirifard:2006wa}
G.~Exirifard, {\it {The alpha-prime stretched horizon in Heterotic string}},
  {\em JHEP} {\bf 10} (2006) 070
  [\href{http://arXiv.org/abs/hep-th/0604021}{{\tt hep-th/0604021}}].

\bibitem{Sen:2007qy}
A.~Sen, {\it {Black Hole Entropy Function, Attractors and Precision Counting of
  Microstates}},  {\em Gen. Rel. Grav.} {\bf 40} (2008) 2249--2431
  [\href{http://arXiv.org/abs/0708.1270}{{\tt 0708.1270}}].

\bibitem{Liu:2008kt}
J.~T. Liu and P.~Szepietowski, {\it {Higher derivative corrections to R-charged
  AdS(5) black holes and field redefinitions}},  {\em Phys. Rev.} {\bf D79}
  (2009) 084042 [\href{http://arXiv.org/abs/0806.1026}{{\tt 0806.1026}}].

\bibitem{Castro:2013pqa}
A.~Castro, N.~Dehmami, G.~Giribet and D.~Kastor, {\it {On the Universality of
  Inner Black Hole Mechanics and Higher Curvature Gravity}},  {\em JHEP} {\bf
  07} (2013) 164 [\href{http://arXiv.org/abs/1304.1696}{{\tt 1304.1696}}].

\bibitem{Mozaffar:2016hmg}
M.~R. Mohammadi~Mozaffar, A.~Mollabashi, M.~M. Sheikh-Jabbari and M.~H.
  Vahidinia, {\it {Holographic Entanglement Entropy, Field Redefinition
  Invariance and Higher Derivative Gravity Theories}},  {\em Phys. Rev.} {\bf
  D94} (2016), no.~4 046002 [\href{http://arXiv.org/abs/1603.05713}{{\tt
  1603.05713}}].

\bibitem{0264-9381-9-5-003}
S.~A. Fulling, R.~C. King, B.~G. Wybourne and C.~J. Cummins, {\it Normal forms
  for tensor polynomials. i. the riemann tensor},  {\em Classical and Quantum
  Gravity} {\bf 9} (1992), no.~5 1151.

\bibitem{Oliva:2010zd}
J.~Oliva and S.~Ray, {\it {Classification of Six Derivative Lagrangians of
  Gravity and Static Spherically Symmetric Solutions}},  {\em Phys. Rev.} {\bf
  D82} (2010) 124030 [\href{http://arXiv.org/abs/1004.0737}{{\tt 1004.0737}}].

\bibitem{Endlich:2017tqa}
S.~Endlich, V.~Gorbenko, J.~Huang and L.~Senatore, {\it {An effective formalism
  for testing extensions to General Relativity with gravitational waves}},
  {\em JHEP} {\bf 09} (2017) 122 [\href{http://arXiv.org/abs/1704.01590}{{\tt
  1704.01590}}].

\bibitem{Deser:2005pc}
S.~Deser and A.~V. Ryzhov, {\it {Curvature invariants of static spherically
  symmetric geometries}},  {\em Class. Quant. Grav.} {\bf 22} (2005) 3315--3324
  [\href{http://arXiv.org/abs/gr-qc/0505039}{{\tt gr-qc/0505039}}].

\bibitem{Ricci}
B.~Andrews and C.~Hopper, {\em {The Ricci Flow in Riemannian Geometry}}.
\newblock Springer, 2011.

\bibitem{Howe:1983sra}
P.~S. Howe and P.~C. West, {\it {The Complete N=2, D=10 Supergravity}},  {\em
  Nucl. Phys.} {\bf B238} (1984) 181--220.

\bibitem{Buchel:2008ae}
A.~Buchel, R.~C. Myers, M.~F. Paulos and A.~Sinha, {\it {Universal holographic
  hydrodynamics at finite coupling}},  {\em Phys. Lett.} {\bf B669} (2008)
  364--370 [\href{http://arXiv.org/abs/0808.1837}{{\tt 0808.1837}}].

\bibitem{Galante:2013wta}
D.~A. Galante and R.~C. Myers, {\it {Holographic Renyi entropies at finite
  coupling}},  {\em JHEP} {\bf 08} (2013) 063
  [\href{http://arXiv.org/abs/1305.7191}{{\tt 1305.7191}}].

\bibitem{Gubser:1998nz}
S.~S. Gubser, I.~R. Klebanov and A.~A. Tseytlin, {\it {Coupling constant
  dependence in the thermodynamics of N=4 supersymmetric Yang-Mills theory}},
  {\em Nucl. Phys.} {\bf B534} (1998) 202--222
  [\href{http://arXiv.org/abs/hep-th/9805156}{{\tt hep-th/9805156}}].

\bibitem{Palais:1979rca}
R.~S. Palais, {\it {The principle of symmetric criticality}},  {\em Commun.
  Math. Phys.} {\bf 69} (1979), no.~1 19--30.

\bibitem{Deser:2003up}
S.~Deser and B.~Tekin, {\it {Shortcuts to high symmetry solutions in
  gravitational theories}},  {\em Class. Quant. Grav.} {\bf 20} (2003)
  4877--4884 [\href{http://arXiv.org/abs/gr-qc/0306114}{{\tt gr-qc/0306114}}].

\bibitem{CHM}
H.~Casini, M.~Huerta and R.~C. Myers, {\it {Towards a derivation of holographic
  entanglement entropy}},  {\em JHEP} {\bf 05} (2011) 036
  [\href{http://arXiv.org/abs/1102.0440}{{\tt 1102.0440}}].

\bibitem{Henningson:1998gx}
M.~Henningson and K.~Skenderis, {\it {The Holographic Weyl anomaly}},  {\em
  JHEP} {\bf 07} (1998) 023 [\href{http://arXiv.org/abs/hep-th/9806087}{{\tt
  hep-th/9806087}}].

\bibitem{Balasubramanian:1999re}
V.~Balasubramanian and P.~Kraus, {\it {A Stress tensor for Anti-de Sitter
  gravity}},  {\em Commun. Math. Phys.} {\bf 208} (1999) 413--428
  [\href{http://arXiv.org/abs/hep-th/9902121}{{\tt hep-th/9902121}}].

\bibitem{Emparan:1999pm}
R.~Emparan, C.~V. Johnson and R.~C. Myers, {\it {Surface terms as counterterms
  in the AdS / CFT correspondence}},  {\em Phys. Rev.} {\bf D60} (1999) 104001
  [\href{http://arXiv.org/abs/hep-th/9903238}{{\tt hep-th/9903238}}].

\bibitem{Bueno:2019ycr}
P.~Bueno, P.~A. Cano and R.~A. Hennigar, {\it {(Generalized) quasi-topological
  gravities at all orders}},  \href{http://arXiv.org/abs/1909.07983}{{\tt
  1909.07983}}.

\end{thebibliography}\endgroup
\bibliographystyle{JHEP-2}
\label{biblio}

\end{document}